\documentclass[aps,prd,preprint,tightenlines,superscriptaddress,showpacs,byrevtex,floatfix]{revtex4}
\usepackage{graphicx} % Include figure files
\usepackage{dcolumn}  % Align table columns on decimal point
\usepackage{bm}
\usepackage{amsmath}
\usepackage{color}
\usepackage{multirow}

\graphicspath{{ps}}

\newcommand{\aonepi}{\ensuremath{B^{0} \to a^{\pm}_{1}\pi^{\mp}}}
\newcommand{\atwopi}{\ensuremath{B^{0} \to a^{\pm}_{2}\pi^{\mp}}}
\newcommand{\aonepilong}{\ensuremath{B^{0} \to a^{\pm}_{1}(1260)\pi^{\mp}}}
\newcommand{\atwopilong}{\ensuremath{B^{0} \to a^{\pm}_{2}(1320)\pi^{\mp}}}
\newcommand{\dpi}{\ensuremath{B^{0} \to D^{-} [K^{+}\pi^{-}\pi^{-}] \pi^{+}}}

\newcommand{\epem}{\ensuremath{e^{+} e^{-}}}
\newcommand{\qqbar}{\ensuremath{q \bar{q}}}
\newcommand{\BBbar}{\ensuremath{B \bar{B}}}
\newcommand{\BzBzb}{\ensuremath{B^{0} \bar{B}^{0}}}
\newcommand{\BpBm}{\ensuremath{B^{+} B^{-}}}

\newcommand{\pip}{\ensuremath{\pi^{+}}}
\newcommand{\pim}{\ensuremath{\pi^{-}}}
\newcommand{\fzsigma}{\ensuremath{f_{0}(600)}}
\newcommand{\rhoz}{\ensuremath{\rho^{0}}}
\newcommand{\fz}{\ensuremath{f_{0}}}
\newcommand{\aone}{\ensuremath{a_{1}^{\pm}}}
\newcommand{\aonep}{\ensuremath{a_{1}^{+}}}
\newcommand{\aonem}{\ensuremath{a_{1}^{-}}}
\newcommand{\atwo}{\ensuremath{a_{2}^{\pm}}}
\newcommand{\bone}{\ensuremath{b_{1}^{\pm}}}
\newcommand{\Ks}{\ensuremath{K^{0}_{S}}}
\newcommand{\Bz}{\ensuremath{B^{0}}}
\newcommand{\Bzb}{\ensuremath{\bar{B}^{0}}}
\newcommand{\Ups}{\ensuremath{\Upsilon(4S)}}

\newcommand{\Brec}{\ensuremath{B^{0}_{\rm Rec}}}
\newcommand{\Btag}{\ensuremath{B^{0}_{\rm Tag}}}

\newcommand{\Mbc}{\ensuremath{M_{\rm bc}}}
\newcommand{\De}{\ensuremath{\Delta E}}
\newcommand{\Fsb}{\ensuremath{{\cal F}_{b \bar b/q \bar q}}}
\newcommand{\mthreepi}{\ensuremath{m_{3\pi}}}
\newcommand{\hel}{\ensuremath{{\cal H}_{3\pi}}}
\newcommand{\Dt}{\ensuremath{\Delta t}}
\newcommand{\Dz}{\ensuremath{\Delta z}}

\newcommand{\taub}{\ensuremath{\tau_{B^{0}}}}
\newcommand{\Dw}{\ensuremath{\Delta w}}
\newcommand{\Dmd}{\ensuremath{\Delta m_{d}}}
\newcommand{\Acp}{\ensuremath{{\cal A}_{CP}}}
\newcommand{\Scp}{\ensuremath{{\cal S}_{CP}}}
\newcommand{\Ccp}{\ensuremath{{\cal C}_{CP}}}
\newcommand{\DS}{\ensuremath{\Delta{\cal S}}}
\newcommand{\DC}{\ensuremath{\Delta{\cal C}}}
\newcommand{\phitwo}{\ensuremath{\phi_{2}}}
\newcommand{\phitwoeff}{\ensuremath{\phi^{\rm eff}_{2}}}

\begin{document}

\vspace*{-3\baselineskip}
%\resizebox{!}{3cm}{\includegraphics{belle.eps}}

\begin{minipage}[]{0.6\columnwidth}
  \includegraphics[height=3.0cm,width=!]{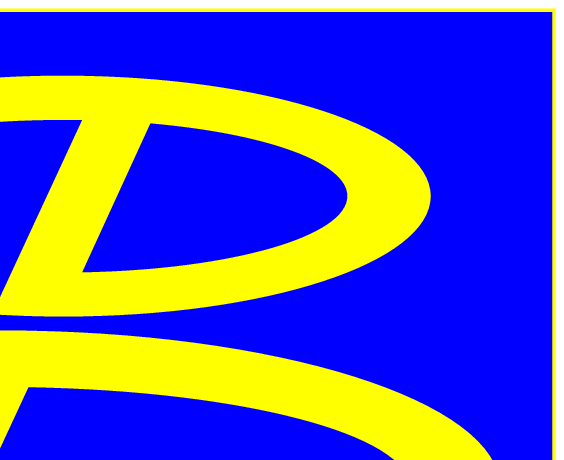}
\end{minipage}
\begin{minipage}[]{0.4\columnwidth}
  %\hbox{Version 1.0.10}
  %\hbox{Intended for Phys. Rev. D}
  %\hbox{Author: J.~Dalseno}
  %\hbox{Committee: P.~Krizan (chair)}
  %\hbox{A.~Garmash, H.~Sahoo}
  \hbox{Belle Preprint 2012-15}
  \hbox{KEK Preprint 2012-7}
\end{minipage}

%\preprint{
%  \vbox{
%    \hbox{}
%    \hbox{Version 1.0}
%    \hbox{Intended for Phys. Rev. D}
%    \hbox{Author: J.~Dalseno}
%    \hbox{Committee: P.~Krizan (chair)}
%    \hbox{A.~Garmash, H.~Sahoo}
    % \hbox{hep-ex nnnn}
%  }
%}

\title{ \quad\\[1.0cm] Measurement of Branching Fraction and First Evidence of\\ $\bm{CP}$ Violation in $\bm{\aonepilong}$ Decays}

%%%% >>>>> insert the authorlist here. BEFORE the abstract !!!!! <<<<<
%%%% >>>>> from the authorship confirmation web page
%%% Name the file author.tex and use \input{author} to insert into your latex file.
%%% Paper:    CPV in B0 -> a_1(1260) pi
%%% Journal:  Physical Review D
%%% Contacts: J. Dalseno (jdalseno@post.kek.jp)
%%% Non-responding authors or those who said NO are commented out.
%%% ====================================================================
%%% Click the RELOAD button on your web browser to see the updated file.
%%% ====================================================================
%%% Use \input{author} to insert this material into your latex file.
%%%%% Force institutions to appear in alphabetical order when typeset.
\affiliation{University of Bonn, Bonn 53115}
\affiliation{Budker Institute of Nuclear Physics SB RAS and Novosibirsk State University, Novosibirsk 630090}
\affiliation{Faculty of Mathematics and Physics, Charles University, 121 16 Prague}
%%%\affiliation{Chiba University, Chiba 263-8522}
\affiliation{University of Cincinnati, Cincinnati, Ohio 45221}
%%%\affiliation{Department of Physics, Fu Jen Catholic University, Taipei 24205}
%%%\affiliation{Justus-Liebig-Universit\"at Gie\ss{}en, 35392 Gie\ss{}en}
\affiliation{Gifu University, Gifu 501-1193}
%%%\affiliation{The Graduate University for Advanced Studies, Hayama 240-0193}
%%%\affiliation{Gyeongsang National University, Chinju 660-701}
\affiliation{Hanyang University, Seoul 133-791}
\affiliation{University of Hawaii, Honolulu, Hawaii 96822}
\affiliation{High Energy Accelerator Research Organization (KEK), Tsukuba 305-0801}
%%%\affiliation{Hiroshima Institute of Technology, Hiroshima 731-5193}
%%%\affiliation{University of Illinois at Urbana-Champaign, Urbana, Illinois 61801}
\affiliation{Indian Institute of Technology Guwahati, Assam 781039}
\affiliation{Indian Institute of Technology Madras, Chennai 600036}
%%%\affiliation{Indiana University, Bloomington, Indiana 47408}
\affiliation{Institute of High Energy Physics, Chinese Academy of Sciences, Beijing 100049}
\affiliation{Institute of High Energy Physics, Vienna 1050}
\affiliation{Institute of High Energy Physics, Protvino 142281}
%%%\affiliation{Institute of Mathematical Sciences, Chennai 600113}
%%%\affiliation{INFN - Sezione di Torino, 10125 Torino}
\affiliation{Institute for Theoretical and Experimental Physics, Moscow 117218}
\affiliation{J. Stefan Institute, 1000 Ljubljana}
\affiliation{Kanagawa University, Yokohama 221-8686}
\affiliation{Institut f\"ur Experimentelle Kernphysik, Karlsruher Institut f\"ur Technologie, Karlsruhe 76131}
\affiliation{Korea Institute of Science and Technology Information, Daejeon 305-806}
\affiliation{Korea University, Seoul 136-713}
%%%\affiliation{Kyoto University, Kyoto 606-8502}
\affiliation{Kyungpook National University, Daegu 702-701}
\affiliation{\'Ecole Polytechnique F\'ed\'erale de Lausanne (EPFL), Lausanne 1015}
\affiliation{Faculty of Mathematics and Physics, University of Ljubljana, 1000 Ljubljana}
%%%\affiliation{Luther College, Decorah, Iowa 52101}
\affiliation{University of Maribor, 2000 Maribor}
\affiliation{Max-Planck-Institut f\"ur Physik, 80805 M\"unchen}
\affiliation{University of Melbourne, School of Physics, Victoria 3010}
\affiliation{Graduate School of Science, Nagoya University, Nagoya 464-8602}
\affiliation{Kobayashi-Maskawa Institute, Nagoya University, Nagoya 464-8602}
%%%\affiliation{Nara University of Education, Nara 630-8528}
\affiliation{Nara Women's University, Nara 630-8506}
\affiliation{National Central University, Chung-li 32054}
\affiliation{National United University, Miao Li 36003}
\affiliation{Department of Physics, National Taiwan University, Taipei 10617}
\affiliation{H. Niewodniczanski Institute of Nuclear Physics, Krakow 31-342}
\affiliation{Nippon Dental University, Niigata 951-8580}
\affiliation{Niigata University, Niigata 950-2181}
%%%\affiliation{University of Nova Gorica, 5000 Nova Gorica}
\affiliation{Osaka City University, Osaka 558-8585}
%%%\affiliation{Osaka University, Osaka 565-0871}
\affiliation{Pacific Northwest National Laboratory, Richland, Washington 99352}
%%%\affiliation{Panjab University, Chandigarh 160014}
%%%\affiliation{Peking University, Beijing 100871}
%%%\affiliation{Princeton University, Princeton, New Jersey 08544}
\affiliation{Research Center for Electron Photon Science, Tohoku University, Sendai 980-8578}
%%%\affiliation{Research Center for Nuclear Physics, Osaka University, Osaka 567-0047}
%%%\affiliation{RIKEN BNL Research Center, Upton, New York 11973}
%%%\affiliation{Saga University, Saga 840-8502}
\affiliation{University of Science and Technology of China, Hefei 230026}
\affiliation{Seoul National University, Seoul 151-742}
%%%\affiliation{Shinshu University, Nagano}
\affiliation{Sungkyunkwan University, Suwon 440-746}
\affiliation{School of Physics, University of Sydney, NSW 2006}
\affiliation{Tata Institute of Fundamental Research, Mumbai 400005}
\affiliation{Excellence Cluster Universe, Technische Universit\"at M\"unchen, 85748 Garching}
\affiliation{Toho University, Funabashi 274-8510}
\affiliation{Tohoku Gakuin University, Tagajo 985-8537}
\affiliation{Tohoku University, Sendai 980-8578}
\affiliation{Department of Physics, University of Tokyo, Tokyo 113-0033}
\affiliation{Tokyo Institute of Technology, Tokyo 152-8550}
\affiliation{Tokyo Metropolitan University, Tokyo 192-0397}
\affiliation{Tokyo University of Agriculture and Technology, Tokyo 184-8588}
%%%\affiliation{Toyama National College of Maritime Technology, Toyama 933-0293}
\affiliation{CNP, Virginia Polytechnic Institute and State University, Blacksburg, Virginia 24061}
%%%\affiliation{Wayne State University, Detroit, Michigan 48202}
\affiliation{Yamagata University, Yamagata 990-8560}
\affiliation{Yonsei University, Seoul 120-749}
  \author{J.~Dalseno}\affiliation{Max-Planck-Institut f\"ur Physik, 80805 M\"unchen}\affiliation{Excellence Cluster Universe, Technische Universit\"at M\"unchen, 85748 Garching} % MPI
  \author{I.~Adachi}\affiliation{High Energy Accelerator Research Organization (KEK), Tsukuba 305-0801} % KEK
% \author{K.~Adamczyk}\affiliation{H. Niewodniczanski Institute of Nuclear Physics, Krakow 31-342} % Krakow
  \author{H.~Aihara}\affiliation{Department of Physics, University of Tokyo, Tokyo 113-0033} % Tokyo
% \author{K.~Arinstein}\affiliation{Budker Institute of Nuclear Physics SB RAS and Novosibirsk State University, Novosibirsk 630090} % BINP
% \author{Y.~Arita}\affiliation{Graduate School of Science, Nagoya University, Nagoya 464-8602} % Nagoya
  \author{D.~M.~Asner}\affiliation{Pacific Northwest National Laboratory, Richland, Washington 99352} % PNNL
% \author{T.~Aso}\affiliation{Toyama National College of Maritime Technology, Toyama 933-0293} % Toyama
  \author{V.~Aulchenko}\affiliation{Budker Institute of Nuclear Physics SB RAS and Novosibirsk State University, Novosibirsk 630090} % BINP
  \author{T.~Aushev}\affiliation{Institute for Theoretical and Experimental Physics, Moscow 117218} % ITEP
% \author{T.~Aziz}\affiliation{Tata Institute of Fundamental Research, Mumbai 400005} % Tata
  \author{A.~M.~Bakich}\affiliation{School of Physics, University of Sydney, NSW 2006} % Sydney
% \author{Y.~Ban}\affiliation{Peking University, Beijing 100871} % Peking
% \author{E.~Barberio}\affiliation{University of Melbourne, School of Physics, Victoria 3010} % Melbourne
% \author{M.~Barrett}\affiliation{University of Hawaii, Honolulu, Hawaii 96822} % Hawaii
  \author{A.~Bay}\affiliation{\'Ecole Polytechnique F\'ed\'erale de Lausanne (EPFL), Lausanne 1015} % Lausanne
% \author{I.~Bedny}\affiliation{Budker Institute of Nuclear Physics SB RAS and Novosibirsk State University, Novosibirsk 630090} % BINP
% \author{M.~Belhorn}\affiliation{University of Cincinnati, Cincinnati, Ohio 45221} % Cincinnati
  \author{K.~Belous}\affiliation{Institute of High Energy Physics, Protvino 142281} % Protvino
% \author{V.~Bhardwaj}\affiliation{Nara Women's University, Nara 630-8506} % Nara
  \author{B.~Bhuyan}\affiliation{Indian Institute of Technology Guwahati, Assam 781039} % IITG
% \author{M.~Bischofberger}\affiliation{Nara Women's University, Nara 630-8506} % Nara
% \author{S.~Blyth}\affiliation{National United University, Miao Li 36003} % NUU
% \author{A.~Bondar}\affiliation{Budker Institute of Nuclear Physics SB RAS and Novosibirsk State University, Novosibirsk 630090} % BINP
% \author{G.~Bonvicini}\affiliation{Wayne State University, Detroit, Michigan 48202} % WayneState
% ...
% Karlsruhe
% \author{N.~Zwahlen}\affiliation{\'Ecole Polytechnique F\'ed\'erale de Lausanne (EPFL), Lausanne 1015} % Lausanne
% \author{O.~Zyukova}\affiliation{Budker Institute of Nuclear Physics SB RAS and Novosibirsk State University, Novosibirsk 630090} % BINP
\collaboration{The Belle Collaboration}
%% end author list

\begin{abstract}
  We present a measurement of the branching fraction and time-dependent $CP$ violation parameters in \aonepilong\ decays. The results are obtained from the final data sample containing $772 \times 10^{6}$ \BBbar\ pairs collected at the \Ups\ resonance with the Belle detector at the KEKB asymmetric-energy \epem\ collider. We obtain the product branching fraction
  \begin{center}
    ${\cal B}(\aonepilong)\times{\cal B}(a_{1}^{\pm}(1260) \rightarrow \pi^{\pm}\pi^{\mp}\pi^{\pm}) = (11.1 \pm 1.0 \; (\rm stat) \pm 1.4 \; (\rm syst) ) \times 10^{-6}$,
  \end{center}
  and an upper limit on the product branching fraction for a possible decay with the same final state
  \begin{center}
    ${\cal B}(\atwopilong)\times{\cal B}(a^{\pm}_{2}(1320) \rightarrow \pi^{\pm}\pi^{\mp}\pi^{\pm}) < 2.2 \times 10^{-6} \textrm{ at 90\% CL}$.
  \end{center}
  In a time-dependent measurement to extract $CP$ asymmetries, we obtain the $CP$ violation parameters
  $$
  \begin{array}{rcl}
    \Acp \!\!\!&=&\!\!\! -0.06 \pm 0.05 \textrm{ (stat)} \pm 0.07 \textrm{ (syst)},\\
    \Ccp \!\!\!&=&\!\!\! -0.01 \pm 0.11 \textrm{ (stat)} \pm 0.09 \textrm{ (syst)},\\
    \Scp \!\!\!&=&\!\!\! -0.51 \pm 0.14 \textrm{ (stat)} \pm 0.08 \textrm{ (syst)},
  \end{array}
  $$
  representing time and flavor integrated direct, flavor-dependent direct and mixing-induced $CP$ violation, respectively. Simultaneously, we also extract the $CP$ conserving parameters
  $$
  \begin{array}{rcl}
    \DC \!\!\!&=&\!\!\! +0.54 \pm 0.11 \textrm{ (stat)} \pm 0.07 \textrm{ (syst)},\\
    \DS \!\!\!&=&\!\!\! -0.09 \pm 0.14 \textrm{ (stat)} \pm 0.06 \textrm{ (syst)},
  \end{array}
  $$
  which, respectively, describe a rate difference and strong phase difference between the decay channels where the \aone\ does not contain the spectator quark and those where it does. We find first evidence of mixing-induced $CP$ violation in \aonepilong\ decays with $3.1\sigma$ significance. The rate where the \aone\ does not contain the spectator quark from the $B$ meson is found to dominate the rate where it does at the $4.1\sigma$ level. However, there is no evidence for either time- and flavor- integrated direct $CP$ violation or flavor-dependent direct $CP$ violation.
\end{abstract}


\maketitle

%%%% >>>> keep the final version single-spaced
\tighten

{\renewcommand{\thefootnote}{\fnsymbol{footnote}}}
\setcounter{footnote}{0}

\section{Introduction}
$CP$ violation in the standard model arises from a complex phase in the Cabibbo-Kobayashi-Maskawa (CKM) quark-mixing matrix~\cite{Cabibbo,KM}. Mixing-induced $CP$ violation in the $B$ sector has been clearly observed by the BaBar~\cite{jpsiks_Belle} and Belle~\cite{jpsiks_BABAR} Collaborations in the $\bar b \rightarrow \bar c c \bar s$ induced decay $\Bz \rightarrow J/\psi \Ks$, while many other modes provide additional information on $CP$ violating parameters.

Decays that proceed dominantly through the $\bar b \rightarrow \bar u u \bar d$ transition are sensitive to the interior angle of the unitarity triangle $\phitwo \, (\alpha) \equiv \arg(-V_{td}V^{*}_{tb})/(V_{ud}V^{*}_{ub})$. The BaBar and Belle Collaborations have reported time-dependent $CP$ asymmetries in these modes that include decays such as $\Bz \rightarrow \pip \pim$~\cite{pipi_Belle,pipi_BABAR}, $\rho^{\pm} \pi^{\mp}$~\cite{rhopi_Belle,rhopi_BABAR} and $\rho^{+} \rho^{-}$~\cite{rhorho_Belle,rhorho_BABAR}.

\begin{figure}
  \centering
  \includegraphics[height=120pt,width=!]{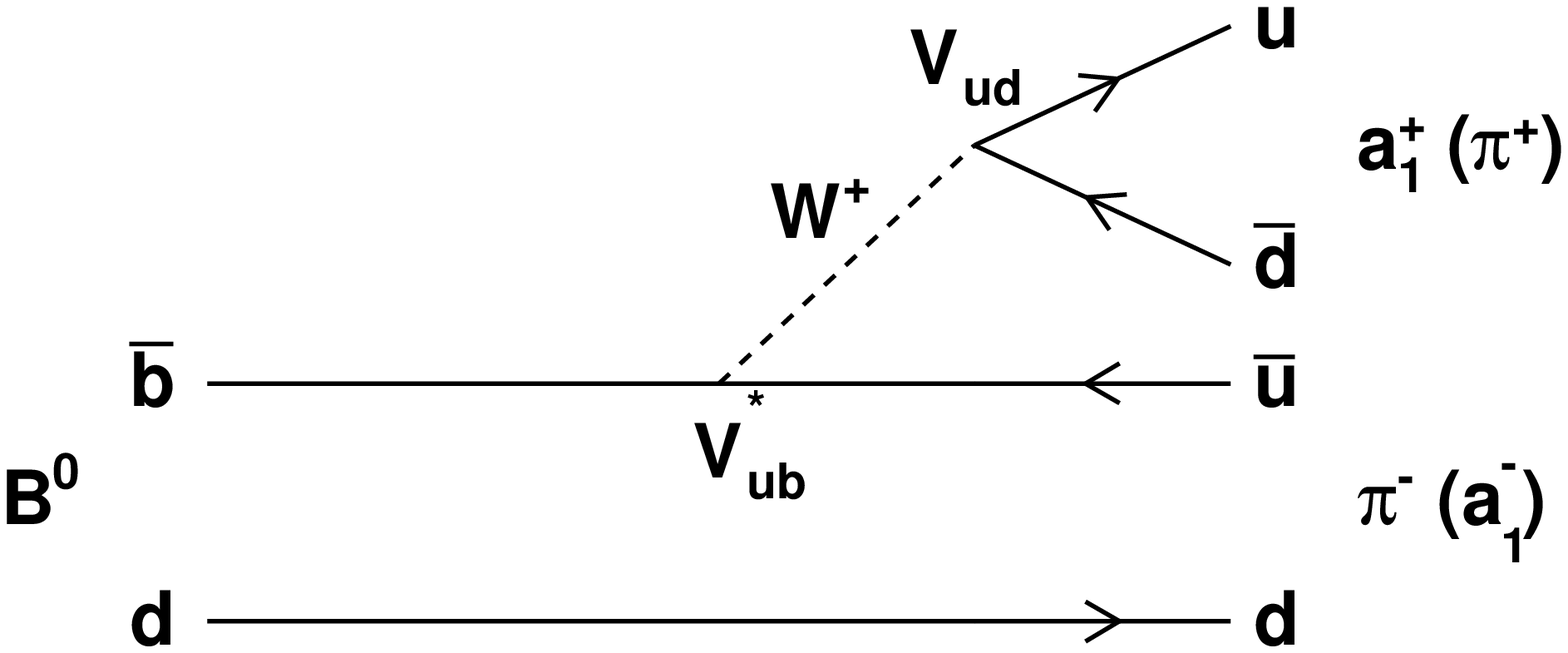}
  \includegraphics[height=120pt,width=!]{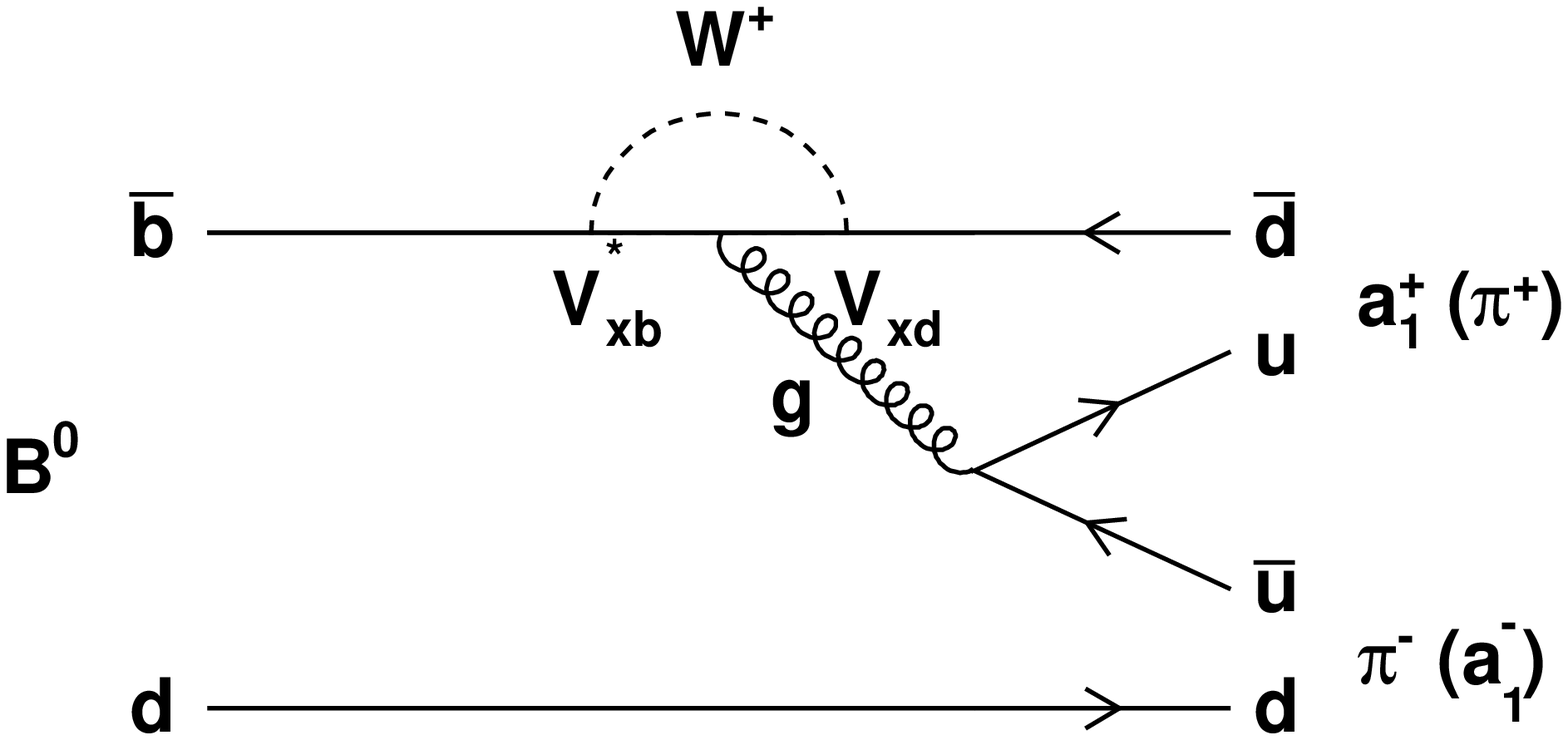}
  \caption{Leading-order tree (left) and penguin (right) diagrams for the decay \aonepi, where the parentheses in the figure indicate the two possible decays of the \Bz. The $d$ quark may become part of the \pim\ or $a_{1}^{-}$. In the penguin diagram, the $x$ in $V_{xb}$ refers to the flavor of the intermediate-state quark $(x=u,c,t)$.}
  \label{fig_a1pi}
\end{figure}

This paper describes the measurement of the branching fraction and time-dependent $CP$ violation parameters of the $\bar b \rightarrow \bar u u \bar d$ channel \aonepi, shown in Fig.~\ref{fig_a1pi}. The left diagram shows the dominant first-order or tree process while the right diagram shows the leading second-order loop or penguin process. This analysis can be used to test the QCD factorization framework which has been used to predict the branching fraction and $CP$ asymmetries of this decay channel~\cite{theory_ap1,theory_ap2,theory_ap3}. Similar to $\rho^{\pm} \pi^{\mp}$, the state $a_{1}^{\pm} \pi^{\mp}$ is not a $CP$ eigenstate; rather, it is a flavor non-specific state with four flavor-charge configurations that must be considered: $\Bz (\Bzb) \rightarrow a_{1}^{\pm} \pi^{\mp}$~\cite{theory_a1pi}. The combined information of the $B$ flavor and $a_1$ charge allows the determination of additional information compared to $CP$ eigenstates. It allows us to separate the cases where the $d$ quark from the $B$ meson, which does not participate in the interaction (spectator), becomes part of the $a^{\pm}_{1}$ or $\pi^{\pm}$.

The decay of the \Ups\ can produce a \BzBzb\ pair that must be coherent, of which one (\Brec) may be reconstructed in the $a^{\pm}_{1} \pi^{\mp}$ final state. This final state does not determine whether the \Brec\ decayed as a \Bz\ or as a \Bzb. The other $B$ meson in the event (\Btag), however, can be reconstructed in a final state that determines its $b$-flavor, and therefore the flavor of the \Brec\ at the time of the \Btag\ decay. The proper time interval between \Brec\ and \Btag, which decay at time $t_{\rm Rec}$ and $t_{\rm Tag}$, respectively, is defined as $\Dt \equiv t_{\rm Rec} - t_{\rm Tag}$. For the case of coherent \BzBzb\ pairs, the time-dependent decay rate in the quasi-two-body approximation when \Btag\ possesses flavor $q$ (\Bz: $q=+1$; \Bzb: $q=-1$) and the $a_{1}$ possesses charge $c$ ($a^{+}_{1}$: $c=+1$; $a^{-}_{1}$: $c=-1$), is given by~\cite{theory_a1pi}
\begin{eqnarray}
  {\cal P}(\Dt, q, c) &=& (1+c\Acp)\frac{e^{-|\Dt|/\taub}}{8\taub} \biggl\{1 + \nonumber \\
  & & q \biggl[(\Scp + c\DS)\sin \Dmd \Dt - (\Ccp + c\DC)\cos \Dmd \Dt\biggr]\biggr\}.
\end{eqnarray}
Here, \taub\ is the \Bz\ lifetime and \Dmd\ is the mass difference between the two mass eigenstates of the neutral $B$ meson. This assumes $CPT$ invariance, no $CP$ violation in the mixing, and that the decay rate difference between the two mass eigenstates is negligible. The parameter \Acp\ measures time and flavor-integrated direct $CP$ violation,
\begin{equation}
  \Acp = \frac{\Gamma(B \to \aonep \pim)-\Gamma(B \to \aonem \pip)}{\Gamma(B \to \aonep \pim)+\Gamma(B \to \aonem \pip)}.
\end{equation}
The parameter \Scp\ measures mixing-induced $CP$ violation, and \Ccp\ measures flavor-dependent direct $CP$ violation. The quantity \DC\ measures the rate asymmetry between the flavor-charge configurations where the $a_1$ does not contain the spectator quark $(\Gamma[\Bz \to a^{+}_{1}\pim] + \Gamma[\Bzb \to a^{-}_{1}\pip])$, and where it does contain the spectator quark $(\Gamma[\Bz \to a^{-}_{1}\pip] + \Gamma[\Bzb \to a^{+}_{1}\pim])$, while \DS\ is related to the strong phase difference between these two processes, $\delta$. These two parameters are not sensitive to $CP$ violation. Sensitivity to \phitwo\ comes from this relation between four of the measured parameters,
\begin{equation}
  \Scp \pm \DS = \sqrt{1 - (\Ccp \pm \DC)^2}\sin(2 \phitwoeff{}^{\pm} \pm \delta).
\end{equation}
A feature common to this channel and the other $\bar b \rightarrow \bar u u \bar d$ modes mentioned earlier is that an effective angle, \phitwoeff, is determined rather than \phitwo\ itself due to the possible presence of additional loop contributions. In the limit that only the dominant tree amplitude contributes, no flavor-dependent $CP$ violation is expected and $\phitwoeff{}^{\pm} = \phitwo$. Fortunately, this inconvenience can be overcome with bounds on $\Delta \phitwo \equiv \phitwo - \phitwoeff$ determined using either an isospin analysis~\cite{theory_isospin} or $SU(3)$ flavor symmetry~\cite{theory_a1pi}.

From these parameters, \phitwoeff\ can be determined up to a four-fold ambiguity~\cite{theory_a1pi},
\begin{equation}
  \phitwoeff = \frac{1}{4}\biggl[ \arcsin \biggl( \frac{\Scp + \DS}{\sqrt{1-(\Ccp + \DC)^{2}}} \biggr) + \arcsin \biggl( \frac{\Scp - \DS}{\sqrt{1-(\Ccp - \DC)^{2}}} \biggr) \biggr].
  \label{eq_phitwo}
\end{equation}
These results can also be transformed into more physically intuitive parameters that characterize direct $CP$ violation in decays with particular topologies~\cite{theory_a1pi2},
\begin{align}
  A_{+-} &= \frac{-(\Acp+\Ccp+\Acp\DC)}{1+\DC+\Acp\Ccp} \nonumber\\
  A_{-+} &= \frac{(-\Acp+\Ccp+\Acp\DC)}{-1+\DC+\Acp\Ccp},
  \label{eq_apm}
\end{align}
which describe $CP$ violation involving diagrams where the $a_1$ contains and does not contain the spectator quark, respectively.

The BaBar Collaboration has performed a branching fraction measurement of \aonepi\ with $218 \,\times\, 10^{6}$ \BBbar\ pairs~\cite{a1pi_BABAR1} and a time-dependent $CP$ violation measurement with $384 \,\times\, 10^{6}$ \BBbar\ pairs~\cite{a1pi_BABAR2}. These results are collected in Table~\ref{tab_a1pi_babar}. From a subsequent study of $B \to K_{1}(1270) \pi$ and $B \to K_{1}(1400) \pi$, the BaBar Collaboration has also obtained bounds on $|\Delta \phitwo|$ using $SU(3)$ flavor symmetry~\cite{a1pi_BABAR3}.
\begin{table}
  \centering
  \caption{Summary of physics parameters for \aonepi\ obtained by the BaBar Collaboration~\cite{a1pi_BABAR1,a1pi_BABAR2}.}
  \begin{tabular}
    {@{\hspace{0.5cm}}c@{\hspace{0.25cm}}  @{\hspace{0.25cm}}c@{\hspace{0.5cm}}}
    \hline \hline
    Parameter & Value\\
    \hline
    ${\cal B}(\aonepi)\times{\cal B}(a_{1}^{\pm}(1260) \rightarrow \pi^{\pm}\pi^{\mp}\pi^{\pm})$ & $(16.6 \pm 1.9 \; ({\rm stat}) \pm 1.5 \; ({\rm syst}) ) \times 10^{-6}$\\
    \Acp & $-0.07 \pm 0.07 \; ({\rm stat}) \pm 0.02 \; ({\rm syst})$\\
    \Ccp & $-0.10 \pm 0.15 \; ({\rm stat}) \pm 0.09 \; ({\rm syst})$\\
    \Scp & $+0.37 \pm 0.21 \; ({\rm stat}) \pm 0.07 \; ({\rm syst})$\\
    \DC & $+0.26 \pm 0.15 \; ({\rm stat}) \pm 0.07 \; ({\rm syst})$\\
    \DS & $-0.14 \pm 0.21 \; ({\rm stat}) \pm 0.06 \; ({\rm syst})$\\
    \hline \hline
  \end{tabular}
  \label{tab_a1pi_babar}
\end{table}

Two separate measurements are described in this paper. In Sec.~\ref{Data Set And Belle Detector}, we briefly describe the data set and Belle detector. The branching fraction measurement is described in Sec.~\ref{Branching Fraction Measurement}. There, we explain the selection criteria used to obtain signal candidates and suppress backgrounds followed by the fit method used to extract the signal component. After this, the results of the fit are presented along with a discussion of the systematic uncertainties. In Sec.~\ref{Time-dependent Measurement}, these same issues are described again for the time-dependent $CP$ violation measurement followed by our conclusions in Sec.~\ref{Conclusion}.

\section{Data Set And Belle Detector}
\label{Data Set And Belle Detector}
This measurement of the branching fraction and time-dependent $CP$ violation parameters in \aonepi\ decays is based on the final data sample containing $772 \times 10^{6}$ \BBbar\ pairs collected with the Belle detector at the KEKB asymmetric-energy \epem\ ($3.5$ on $8~{\rm GeV}$) collider~\cite{KEKB}. At the \Ups\ resonance ($\sqrt{s}=10.58$~GeV), the Lorentz boost of the produced \BBbar\ pairs was $\beta\gamma=0.425$ along the $z$ direction, which is opposite the positron beam direction.

% {\it SVD1:}
% {The Belle detector is a large-solid-angle magnetic
% {spectrometer that
% {consists of a three-layer silicon vertex detector (SVD),
% {a 50-layer central drift chamber (CDC), an array of
% {aerogel threshold Cherenkov counters (ACC), % <- \v{C}erenkov 2007.08
% {a barrel-like arrangement of time-of-flight
% {scintillation counters (TOF), and an electromagnetic calorimeter
% {comprised of CsI(Tl) crystals (ECL) located inside 
% {a super-conducting solenoid coil that provides a 1.5~T
% {magnetic field.  An iron flux-return located outside of
% {the coil is instrumented to detect $K_L^0$ mesons and to identify
% {muons (KLM).  The detector
% {is described in detail elsewhere~\cite{Belle}.

% {\it SVD2+SVD1:}
The Belle detector is a large-solid-angle magnetic
spectrometer that consists of a silicon vertex detector (SVD),
a 50-layer central drift chamber (CDC), an array of
aerogel threshold Cherenkov counters (ACC),  % <- \v{C}erenkov 2007.08
a barrel-like arrangement of time-of-flight
scintillation counters (TOF), and an electromagnetic calorimeter
comprising of CsI(Tl) crystals (ECL) located inside 
a superconducting solenoid coil that provides a 1.5~T
magnetic field.  An iron flux-return located outside of
the coil is instrumented to detect $K_L^0$ mesons and to identify
muons (KLM).  The detector
is described in detail elsewhere~\cite{Belle}.
% {\bf SVD2+SVD1, up to experiment 37:}
Two inner detector configurations were used. A 2.0 cm radius beampipe
and a 3-layer silicon vertex detector (SVD1) was used for the first sample
of $152 \times 10^6 B\bar{B}$ pairs, while a 1.5 cm radius beampipe, a 4-layer
silicon detector (SVD2) and a small-cell inner drift chamber were used to record  
the remaining $620 \times 10^6 B\bar{B}$ pairs~\cite{svd2}. We use a GEANT-based
 Monte Carlo (MC) simulation to model the response of the detector and determine
 its acceptance~\cite{GEANT}.

\section{Branching Fraction Measurement}
\label{Branching Fraction Measurement}
\subsection{Event Selection}
We reconstruct \aonepi\ where $\aone \rightarrow (\pip \pim) \pi^{\pm}$. The \aone\ decay proceeds mainly through the $\rhoz \pi^{\pm}$ and $\fzsigma \pi^{\pm}$ intermediate states~\cite{PDG}. We assume that the $\rhoz \pi^{\pm}$ intermediate state gives the dominant contribution and treat the $\fzsigma \pi^{\pm}$ contribution separately in the systematic uncertainties. Thus, the signal MC events for establishing the selection criteria are generated as \aonepi\ decays where $\aone \rightarrow \rhoz \pi^{\pm}$. The \aone\ mass and width are taken to be $m_{a_{1}} = 1.23 \; {\rm GeV}/c^{2}$ and $\Gamma_{a_{1}} = 0.40 \; {\rm GeV}/c^{2}$~\cite{PDG}.

Charged tracks are identified using a loose requirement of distance of closest approach to the interaction point (IP) along the beam direction, $|dz| < 4.0 \; {\rm cm}$, and in the transverse direction, $dr < 0.4 \; {\rm cm}$. With information obtained from the CDC, ACC and TOF, particle identification (PID) is determined with the likelihood ratio ${\cal L}_{i}/({\cal L}_{i} + {\cal L}_{j})$. Here, ${\cal L}_{i}$ (${\cal L}_{j}$) is the likelihood that the particle is of type $i$ ($j$). A requirement of ${\cal L}_{K/\pi} < 0.4$ is placed on all charged pion candidates, which retains 91\% of all pions from \aonepi, but only 9\% of kaons. To further suppress background from particle misidentification, vetoes are applied on particles consistent with the electron or proton hypotheses. Additional SVD requirements of two $z$ hits and one $r-\phi$ hit~\cite{ResFunc} are imposed on the charged tracks so that a good quality vertex of the reconstructed $B$ candidate can be determined.

The intermediate dipion state is reconstructed above the \Ks\ region with an invariant mass $0.52 \; {\rm GeV}/c^{2} < m(\pip \pim) < 1.1 \; {\rm GeV}/c^{2}$. This is combined with another pion and forms an \aone\ candidate if the invariant mass is in the window $0.85 \; {\rm GeV}/c^{2} < \mthreepi < 1.75 \; {\rm GeV}/c^{2}$, below the charm threshold. Upon combination with another pion, a $B$ candidate is formed.

Reconstructed $B$ candidates are described with two nearly uncorrleated kinematic variables: the beam-energy-constrained mass $\Mbc \equiv \sqrt{(E^{\rm CMS}_{\rm beam})^{2} - (p^{\rm CMS}_{B})^{2}}$, and the energy difference $\De \equiv E^{\rm CMS}_{B} - E^{\rm CMS}_{\rm beam}$, where $E^{\rm CMS}_{\rm beam}$ is the beam energy and $E^{\rm CMS}_{B}$ ($p^{\rm CMS}_{B}$) is the energy (momentum) of the $B$ meson, all evaluated in the center-of-mass system (CMS). The $B$ candidates that satisfy $\Mbc > 5.27 \; {\rm GeV}/c^{2}$ and $|\De| < 0.1 \; {\rm GeV}$ are selected for further analysis.

To reduce combinatorial background in forming the \aone\ candidate, the cosine of the angle between the prompt pion from $\aone \rightarrow \rhoz \pi^{\pm}$ and the $B$ candidate in the \aone\ rest frame is required to be between $-0.85$ and $+0.85$. This distribution is roughly flat for signal while peaking at $\pm 1$ for combinatorial background. This selection retains 80\% of signal events while rejecting 43\% of the combinatorial background.

\begin{figure}
  \centering
  \includegraphics[height=300pt,width=!]{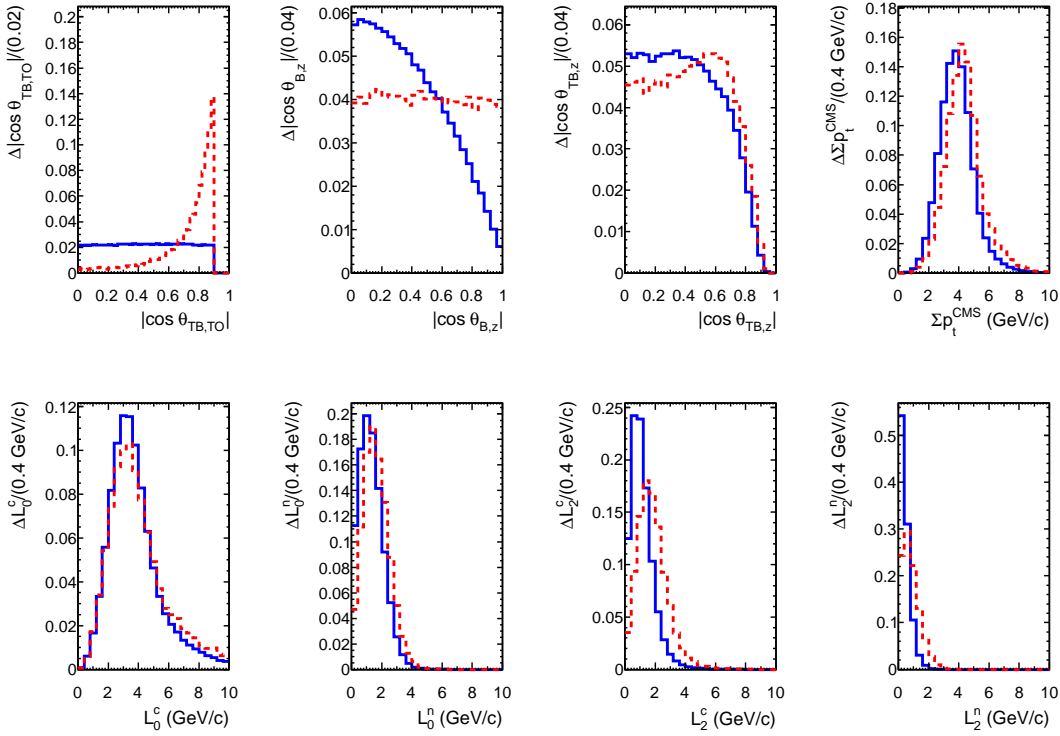}
  \caption{(color online) Simulated (MC) and off-resonance data distributions for the quantities used to construct the Fisher discriminant \Fsb, normalized to have the same area. The solid blue histograms show signal MC events while the dashed red histograms show continuum events from data taken below the \Ups\ resonance.}
  \label{fig_fd}
\end{figure}

The dominant background in the reconstruction of \Brec\ is from continuum ($\epem \rightarrow \qqbar$ where $q=u,d,s,c$) events. Since their topology tends to be jet-like, in contrast to the spherical \BBbar\ decay, continuum events can be distinguished from \BBbar\ events using event-shape variables, which we combine into a Fisher discriminant \Fsb~\cite{nazi_stuff}. The \BBbar\ training sample is taken from signal MC, while the \qqbar\ training sample comes from data taken below the \Ups\ resonance. The Fisher discriminant is then constructed from the following 8 variables:
\begin{itemize}
\item $|\cos \theta_{\rm TB, TO}|$, where the angle is between the \Brec\ thrust direction and the thrust of the tag side. The thrust is defined as the vector which maximizes the sum of the longitudinal momenta of the particles. For a \BBbar\ event, the pair is essentially at rest in the CMS, so the thrust axis of \Brec\ is uncorrelated with the thrust axis of \Btag. In a \qqbar\ event, the decay products lie in the two jets which are back-to-back, so the two thrust axes tend to be collinear. This variable provides the strongest discrimination against continuum.
\item $|\cos \theta_{B, z}|$, where the angle is between the \Brec\ flight direction and the $z$ direction. This is the second most important variable in discriminating against continuum. \BBbar\ pairs are produced in a correlated state, for which this variable follows a sine squared distribution, whereas for \qqbar\ events, the distribution is uniform.
\item $|\cos \theta_{{\rm TB}, z}|$, where the angle is between the \Brec thrust direction and the $z$ direction. This distribution is strongly influenced by detector acceptance at large values. As thrust and decay axes are related, this variable displays similar tendencies to the previously described variable.
\item $\sum p^{\rm CMS}_{t}$, where the transverse CMS momentum sum runs over all particles on the tag side. The continuum distribution generally has a higher mean because its decay product multiplicity is lower compared to \BBbar.
\item $L^{\rm c,n}_{0} \equiv \sum_{c,n} p^{\rm CMS}_{c,n}$, where the CMS momentum sum runs over either the charged tracks ($c$) or neutral clusters ($n$) on the tag side. These distributions exhibit similar tendencies to the previous variable. %Continuum events tend to have a larger mean value because reconstructed $B$ candidates tend to populate the negative \De\ region. Therefore, conservation of the CMS momentum means that the tracks or neutral clusters on the tag side should tend to have larger momentum.
\item $L^{\rm c,n}_{2} \equiv \sum_{c,n} p^{\rm CMS}_{c,n}\cos^{2} \theta_{p^{\rm CMS}_{c,n},{\rm TB}}$, where the CMS momentum sum runs over either the charged tracks ($c$) or neutral clusters ($n$) on the tag side. The angle is between the particle direction and the \Brec\ thrust direction. In addition to the factors explained for the previous variable, the mean of the continuum distribution increases all the more due to higher values of $\cos \theta_{p^{\rm CMS}_{c,n},{\rm TB}}$. For jet-like events like continuum, the momentum of a particle in \Btag\ is closely aligned with its thrust, which itself is strongly correlated with the \Brec\ thrust as explained for the first variable.
\end{itemize}
The distributions for each of these discriminating variables are shown for simulated (MC) signal and continuum (data) events in Fig.~\ref{fig_fd}. Before training, a loose requirement of $\cos \theta_{\rm TB, TO} < 0.9$ is placed that retains 90\% of the signal while rejecting 50\% of the continuum background. The Fisher discriminant is also required to satisfy $-3 < \Fsb < 2$.

The next largest background comes from charm ($b \to c$) and charmless ($b \to u,d,s$) decays of the $B$ meson and is found to exhibit peaking structure in the signal region due to the reconstruction of particular channels with a four-track final state. Defining the decay chain $\Bz \to a_1 \pi_1$, $a_1 \to \rho \pi_2$, $\rho \to \pi_3 \pi_4$, we apply vetoes to remove these peaking backgrounds as summarized in Table~\ref{tab_veto}.
\begin{table}
  \centering
  \caption{Summary of peaking background vetoes. Alternate mass hypotheses have been applied to specific tracks where indicated in order to remove certain channels. The efficiency loss caused by these vetoes is also included. The $X$ represents any charged track(s) that lead to a 4-body final state.}
  \begin{tabular}
    {@{\hspace{0.5cm}}c@{\hspace{0.25cm}}  @{\hspace{0.25cm}}c@{\hspace{0.25cm}}  @{\hspace{0.25cm}}c@{\hspace{0.5cm}}}
    \hline \hline
    Regions vetoed & Modes vetoed & Efficiency loss\\
    \hline
    $1.85 \; {\rm GeV}/c^{2} < m(\pi_{2}K_{3}\pi_{4}) < 1.89 \; {\rm GeV}/c^{2}$ & \multirow{2}{*}{$B \rightarrow D^{+} [K^{-} \pip \pip] X$} & \multirow{2}{*}{$3.7\%$}\\
    $1.85 \; {\rm GeV}/c^{2} < m(\pi_{2}\pi_{3}K_{4}) < 1.89 \; {\rm GeV}/c^{2}$ & &\\
    \hline
    $3.06 \; {\rm GeV}/c^{2} < m(\mu_{1}\mu_{2}) < 3.14 \; {\rm GeV}/c^{2}$ & \multirow{3}{*}{$B \rightarrow J/\psi [\mu^{+} \mu^{-}] X$} & \multirow{3}{*}{$6.1\%$}\\
    $3.06 \; {\rm GeV}/c^{2} < m(\mu_{1}\mu_{3}) < 3.14 \; {\rm GeV}/c^{2}$ & &\\
    $3.06 \; {\rm GeV}/c^{2} < m(\mu_{1}\mu_{4}) < 3.14 \; {\rm GeV}/c^{2}$ & &\\
    \hline
    $0.480 \; {\rm GeV}/c^{2} < m(\pi_{2}\pi_{3}) < 0.516 \; {\rm GeV}/c^{2}$ & \multirow{2}{*}{$B \rightarrow \Ks [\pip \pim] X$} & \multirow{2}{*}{$11.9\%$}\\
    $0.480 \; {\rm GeV}/c^{2} < m(\pi_{2}\pi_{4}) < 0.516 \; {\rm GeV}/c^{2}$ & &\\
    \hline \hline
  \end{tabular}
  \label{tab_veto}
\end{table}

The background coming from $b \rightarrow u \bar u d$ channels with the same final state as signal is studied separately. We consider a possible contribution from \atwopi\ by constructing the helicity variable \hel, defined as the cosine of the angle between the normal to the plane of the $a_1$ candidate and the flight direction of the bachelor pion from the $B$ evaluated in the $3\pi$ rest frame.

On average, 1.6 $B$ candidates are reconstructed per signal event. Selecting the best $B$ candidate having the nearest \Mbc\ with respect to the nominal $B$ meson mass, the correct $B$ is chosen in $91\%$ of cases where the event contains multiple candidates. The fraction of mis-reconstructed signal events in general is $17\%$. As this procedure introduces bias to the \Mbc\ distribution, this variable is excluded from the fit to extract the signal yield.

The full selection used to define the event sample for the $CP$ asymmetry measurement is also applied for the branching fraction determination. Since the \Brec\ and \Btag\ mesons are approximately at rest in the \Ups\ CMS, the difference in decay time between the two $B$ mesons, $\Delta t$, can be approximately determined from the displacement in $z$ between the final state decay vertices,
\begin{equation}
  \Dt \simeq \frac{(z_{\rm Rec} - z_{\rm Tag})}{\beta \gamma c} \equiv \frac{\Delta z}{\beta \gamma c}.
\end{equation}

The vertex of reconstructed $B$ candidates is determined from the charged daughters using the known IP. The IP profile is smeared in the plane perpendicular to $z$ to account for the finite flight length of the $B$ meson in this plane. To obtain the \Dt\ distribution, we reconstruct the tag side vertex from the tracks not used to reconstruct \Brec~\cite{ResFunc}. The events must satisfy the requirements $|\Dt| < 70 \; {\rm ps}$ and $h_{\rm Rec, Tag} < 500$, where $h_{\rm Rec, Tag}$ is the multi-track vertex goodness-of-fit, calculated in three-dimensional space without using the interaction-region profile's constraint.  To reduce the necessity of also modelling the event-dependent observables describing the variable \Dt\ resolution in the fit, the vertex uncertainty is required to be $\sigma^{\rm Rec, Tag}_z < 200 \; \mu {\rm m}$ for multi-track vertices and $\sigma^{\rm Rec, Tag}_z < 500 \; \mu {\rm m}$ for single-track vertices.

After these selection criteria, the MC detection efficiencies are found to be
\begin{eqnarray}
  \label{eq_eff}
  {\rm SVD1:} \; \epsilon(\aonepi) &=& 0.1713 \pm 0.0004 \nonumber \\
  {\rm SVD2:} \; \epsilon(\aonepi) &=& 0.2037 \pm 0.0005,
\end{eqnarray}
where the uncertainty comes from limited MC statistics. Using independent data samples, we also determine correction factors to these efficiencies that account for the difference between data and MC. Correction factors in our reconstruction algorithm arise only from differences in PID and are determined from an inclusive $D^{*+} \to D^{0} [K^{-} \pip] \pip$ sample to be
\begin{eqnarray}
  \label{eq_cf}
  {\rm SVD1:} \; \eta(\aonepi) &=& 0.860 \pm 0.031 \nonumber \\
  {\rm SVD2:} \; \eta(\aonepi) &=& 0.855 \pm 0.047.
\end{eqnarray}

We employ the flavor tagging routine described in Ref.~\cite{Tagging}. The tagging information is represented by two parameters, the \Btag\ flavor $q$ and the purity $r$. The parameter $r$ is an event-by-event MC determined flavor-tagging parameter that ranges from $r = 0$ for no flavor discrimination to $r = 1$ for unambiguous flavor assignment. Due to a finite mistag probability $w$, the $CP$ asymmetry is diluted by a factor $1-2w$. The measure of the performance of the flavor tagging algorithm is the total effective tagging efficiency $\epsilon_{\rm eff} = \epsilon_{\rm Tag}(1-2w)^2$, as the statistical significance of the $CP$ parameters is proportional to $(1-2w)\sqrt{\epsilon_{\rm Tag}}$, where $\epsilon_{\rm Tag}$ is the raw tagging efficiency. These are determined to be $\epsilon_{\rm eff} = 0.284\pm0.010$ and $\epsilon_{\rm eff} = 0.301\pm0.004$ for SVD1 and SVD2 data, respectively.

\subsection{Event Model}
The branching fraction is extracted from a four-dimensional extended unbinned maximum likelihood fit to \De, \Fsb, \mthreepi\ and \hel\ from a data sample divided into 7 bins ($l = 0..6$) in the flavor-tag quality $r$ and 2 SVD configurations $s$. We consider 12 categories in the event model: correctly reconstructed signal (referred to as truth signal hereafter), mis-reconstructed signal, continuum, charm neutral and charged \BBbar\ decays, charmless neutral and charged \BBbar\ decays, and five peaking backgrounds. In most categories, the linear correlations between fit variables are small, so the probability density function (PDF) for each category $j$, is taken as the product of individual PDFs for each variable ${\cal P}^{l,s}_{j}(\De,\Fsb,\mthreepi,\hel) = {\cal P}^{l,s}(\De) \times {\cal P}^{l,s}(\Fsb) \times {\cal P}^{l,s}(\mthreepi) \times {\cal P}^{l,s}(\hel)$ in each $l,s$ bin, unless otherwise mentioned.

The truth model shape is determined from correctly reconstructed signal MC events. The PDF for \De\ is taken to be the sum of two asymmetric-width (bifurcated) Gaussians incorporating calibration factors that correct for the difference between data and MC. These factors calibrate the mean and width of the core bifurcated Gaussian and are determined from a large-statistics control sample \dpi. The PDF for \Fsb\ here and throughout this analysis, for all 12 categories, is the sum of two bifurcated Gaussians in each flavor-tag bin $l$. The shape for all \BBbar\ categories is fixed from the truth model except for the mean of the core distribution, incorporating calibration factors that correct for the shape difference between data and MC. The \mthreepi\ distribution is modelled with an efficiency-corrected relativistic Breit-Wigner
\begin{equation}
  {\cal P}^{l,s}_{\rm Sig}(\mthreepi) \equiv \epsilon^{s}(\mthreepi)\frac{m_{a_{1}}\Gamma(\mthreepi)}{(\mthreepi^{2}-m_{a_{1}}^{2})^{2} + m_{a_{1}}^{2}\Gamma^{2}(\mthreepi)},
\end{equation}
where $\epsilon^{s}$ is the mass-dependent detection efficiency and $\Gamma$ is the mass-dependent width
\begin{equation}
  \Gamma(m) = \Gamma_{a_{1}} \frac{\rho^{1+S}(\mthreepi)}{\rho^{1+S}(m_{a_{1}})}.
\end{equation}
Due to the finite width of the \rhoz, the phase space $\rho^{1+S}$ of the \aone\ decay, where the superscript represents the decay into a spin-1 meson in an $S$-wave configuration~\cite{aone}, cannot be calculated simply. Therefore, we model the phase space empirically with a 6th order Chebyshev polynomial
\begin{equation}
  \rho^{1+S}(\mthreepi) \equiv 1+\sum_{i=1}^{6}c_{i}C_{i}(\mthreepi),
\end{equation}
where $C_{i}$ is a Chebyshev polynomial of order $i$ and $c_{i}$ is the fit coefficient. The PDF for \hel\ is a sum of symmetric Chebyshev polynomials up to 6th order.

The mis-reconstructed model shape is determined from incorrectly reconstructed signal MC events. For \De, the PDF is taken as a smoothed one-dimensional histogram, the \mthreepi\ PDF is the sum of two asymmetric-width Gaussians and the PDF for \hel\ is the sum of symmetric Chebyshev polynomials up to 8th order.

The parameterization of the continuum model is chosen based on data taken below the \Ups\ resonance; however, all the shape parameters are free in the fit to extract the branching fraction. Since continuum is by far the dominant component, extra care must be taken to ensure this background shape is understood, so correlations above 2\% are considered. The PDF for \De\ is taken to be a 1st order Chebyshev polynomial in each flavor-tag bin with a coefficient depending linearly on \Fsb
\begin{equation}
  {\cal P}^{l,s}_{\qqbar}(\De|\Fsb) = 1 + (p^{l,s}_{0} + p^{s}_{1}\Fsb)\De.
\end{equation}
The \mthreepi\ shape is observed to shift quadratically with \Fsb, so the PDF is a sum of Chebyshev polynomial up to 4th order that incorporates an offset. In addition, a small excess is seen above this distribution and is modelled with a Gaussian. The result is
\begin{equation}
  {\cal P}^{l,s}_{\qqbar}(\mthreepi|\Fsb) \equiv (1-f^{s})\biggl[1+\sum_{k=1}^{4}c^{s}_{k}C_{k}(\mthreepi - p^{s}_{1}\Fsb - p^{s}_{2}\Fsb^{2})\biggr] + f^{s}G(\mthreepi; \mu^{s}, \sigma^{s}).
\end{equation}
The PDF for \hel\ is the sum of symmetric Chebyshev polynomials up to 6th order and a mirrored Gaussian
\begin{equation}
  {\cal P}^{l,s}_{\qqbar}(\hel) \equiv (1-2f^{s})\biggl[1+\sum_{k=1}^{6}c^{s}_{k}C_{k}(\hel)\biggr] + f^{s}G(\hel; \mu^{s}, \sigma^{s}) + f^{s}G(\hel; -\mu^{s}, \sigma^{s}).
\end{equation}

The charm \BBbar\ background shape is determined from a large sample of MC containing generic $b \to c$ transitions and is subdivided into neutral and charged $B$ samples. For \De, the PDF is a smoothed one-dimensional histogram; for \mthreepi, the PDF is the sum of Chebyshev polynomials up to 4th order; for \hel, the PDF is the sum of symmetric Chebyshev polynomials up to 6th order plus a mirrored Gaussian.

The charmless \BBbar\ background shape is determined from a large sample of MC containing generic $b \to u,d,s$ transitions and is subdivided into neutral and charged $B$ samples. A sizeable correlation is seen between \De\ and \mthreepi\ and is taken into account with a smoothed two-dimensional histogram. The \hel\ PDF is the sum of symmetric Chebyshev polynomials up to 6th order with a mirrored Gaussian.

Many decay channels contain the same final state as \aonepi. In addition to the possibility of \atwopi, we also consider those listed in Table~\ref{tab_pb} which includes their expected yields in $772 \times 10^{6}$ \BBbar\ pairs. With the exception of \atwopi, the assumed peaking background branching fractions are fixed in the nominal fit from the Heavy Flavor Averaging Group~\cite{HPoofter}. Where a mode is to be included in the fit model but only an upper limit is known, the branching fraction is taken as half the upper limit unless the mode contains an \fz, in which case it is assigned zero branching fraction and instead considered solely in the systematic uncertainties.
\begin{table}
  \centering
  \caption{List of peaking backgrounds, assumed branching fractions and their respective expected yields after all selection criteria has been applied.}
  \begin{tabular}
    {@{\hspace{0.5cm}}c@{\hspace{0.25cm}}  @{\hspace{0.25cm}}c@{\hspace{0.25cm}}  @{\hspace{0.25cm}}c@{\hspace{0.25cm}}  @{\hspace{0.25cm}}c@{\hspace{0.25cm}}  @{\hspace{0.25cm}}c@{\hspace{0.5cm}}}
    \hline \hline
    Mode & ${\cal B}$ $(\times 10^{-6})$ & SVD1 expected events & SVD2 expected events\\
    \hline
    $\Bz \rightarrow \rhoz \rhoz$ & $0.73 \pm 0.28$ & $2$ & $11$\\
    $\Bz \rightarrow \bone [\pi^{\pm} \pi^{\mp} \pi^{\pm}] \pi^{\mp}$ & $0.16 \pm 0.03$ & $5$ & $23$\\
    $\Bz \rightarrow \rhoz \pip \pim$ & $4.35 \pm 4.35$ & $28$ & $137$\\
    $\Bz \rightarrow \pip \pim \pip \pim$ & $9.65 \pm 9.65$ & $26$ & $125$\\
    $\Bz \rightarrow \fz \rhoz$ & $<0.3$ & $<2$ & $<7$\\
    $\Bz \rightarrow \fz \fz$ & $<0.1$ & $<1$ & $<1$\\
    $\Bz \rightarrow \fz \pip \pim$ & $<3.8$ & $<14$ & $<65$\\
    \hline \hline
  \end{tabular}
  \label{tab_pb}
\end{table}
The peaking background shapes are determined from individually generated MC samples and are fixed in the fit to data. The PDF for \De\ borrows the shape of correctly reconstructed signal events and includes a 1st order Chebyshev polynomial to model the mis-reconstructed contribution underneath. The \mthreepi\ PDF depends on the peaking background. For \atwopi, a sum of three Gaussians is used; for $\Bz \rightarrow \rhoz \rhoz$ and $\bone \pi^{\mp}$, a smoothed histogram is used; otherwise, the sum of Chebyshev polynomials up to 4th order is used. The \hel\ PDF is the sum of symmetric Chebyshev polynomials up to 8th order, except for $\Bz \rightarrow \rhoz \rhoz$, which is modelled with a smoothed symmetrized one-dimensional histogram.

The total likelihood for $208238$ \aonepi\ candidates in the fit region is
\begin{equation}
  {\cal L} \equiv \prod_{l,s} \frac{e^{-\sum_{j}N^{s}_{j}\sum_{l,s}f^{l,s}_{j}}}{N_{l,s}!} \prod^{N_{l,s}}_{i=1} \sum_{j}N^{s}_{j}f^{l,s}_{j}{\cal P}^{l,s}_{j}(\De^{i},\Fsb^{i},\mthreepi^{i},\hel^{i}),
\end{equation}
which iterates over $i$ events, $j$ categories, $l$ flavor-tag bins and $s$ detector configurations. Instead of two free signal yields $N^{s}_{\rm Sig}$, the branching fraction is chosen as a single free parameter and is incorporated into the fit with
\begin{equation}
  N^{s}_{\rm sig} = {\cal B}(\aonepi)\times{\cal B}(\aone \rightarrow \pi^{\pm}\pi^{\mp}\pi^{\pm})N^{s}_{\BBbar}\epsilon^{s}_{\rm Sig}\eta^{s}_{\rm Sig},
\end{equation}
where $\epsilon^{s}_{\rm Sig}$ and $\eta^{s}_{\rm Sig}$ are given in Eqs.~\ref{eq_eff}~and~\ref{eq_cf}, respectively. Similar conversions are done for the remaining peaking backgrounds using their expected values from Table~\ref{tab_pb}. The fraction of events in each flavor-tag bin $l$, for category $j$, is denoted by $f^{l,s}_{j}$. The fraction of signal events in each $l,s$ bin, $f^{l,s}_{\rm Sig}$, has been calibrated with the \dpi\ control sample. Other free physics parameters include the \aone\ width and the product branching fraction ${\cal B}({\atwopi})\times{\cal B}(a_{2}^{\pm} \rightarrow \pi^{\pm}\pi^{\mp}\pi^{\pm})$. Also free are the yields $N^{s}_{\qqbar}$, $N^{{\rm charm}; \, s}_{\BzBzb}$ and $N^{{\rm charmless}; \, s}_{\BzBzb}$; the remaining yields are fixed to the values given in Table~\ref{tab_bf_fixed} as determined from MC. In total, there are 121 free parameters in the fit.
\begin{table}
  \centering
  \caption{Summary of yields fixed relative to other yields free in the fit where the uncertainties are from limited MC statistics. The mis-reconstructed yield is fixed relative to the signal yield, and the charm and charmless \BpBm\ background yields are fixed relative to their respective \BzBzb\ background yields.}
  \begin{tabular}
    {@{\hspace{0.5cm}}c@{\hspace{0.25cm}}  @{\hspace{0.25cm}}c@{\hspace{0.25cm}}  @{\hspace{0.25cm}}c@{\hspace{0.5cm}}}
    \hline \hline
    Yield & SVD1 & SVD2\\
    \hline
    $N^{s}_{\rm Mis}$ & $(0.167 \pm 0.001)N^{\rm SVD1}_{\rm Sig}$ & $(0.166 \pm 0.001)N^{\rm SVD2}_{\rm Sig}$\\
    $N^{{\rm charm};\,s}_{\BpBm}$ & $(1.585 \pm 0.019)N^{\rm charm; \, SVD1}_{\BzBzb}$ & $(1.700 \pm 0.010)N^{\rm charm; \, SVD2}_{\BzBzb}$\\
    $N^{{\rm charmless};\,s}_{\BpBm}$ & $(0.568 \pm 0.006)N^{\rm charmless; \, SVD1}_{\BzBzb}$ & $(0.574 \pm 0.003)N^{\rm charmless; \, SVD2}_{\BzBzb}$\\
    \hline \hline
  \end{tabular}
  \label{tab_bf_fixed}
\end{table}
In addition, all shape parameters of the continuum model are free in the fit to data.

\subsection{Fit Result}
\begin{figure}
  \centering
  \includegraphics[height=220pt,width=!]{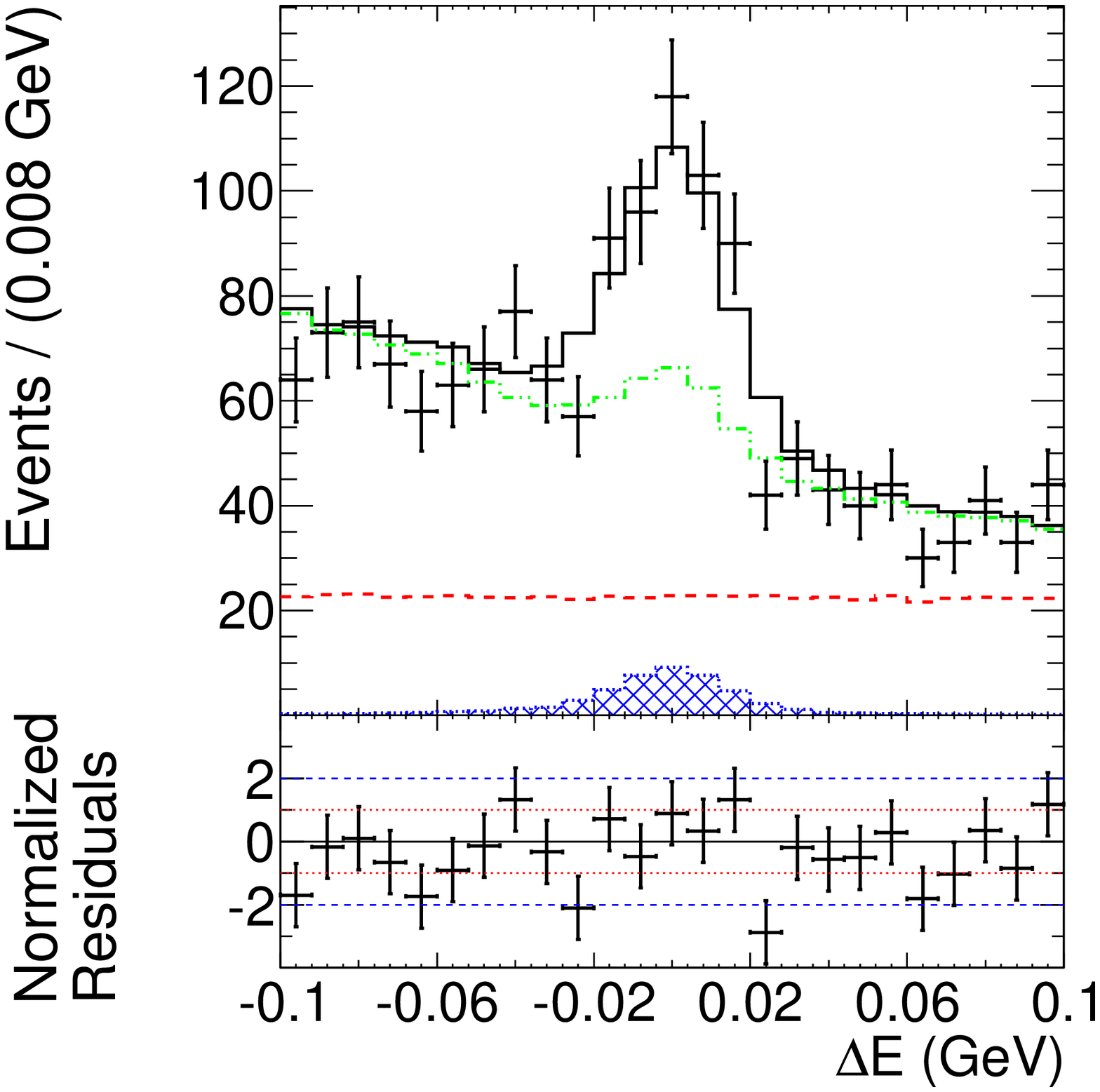}
  \includegraphics[height=220pt,width=!]{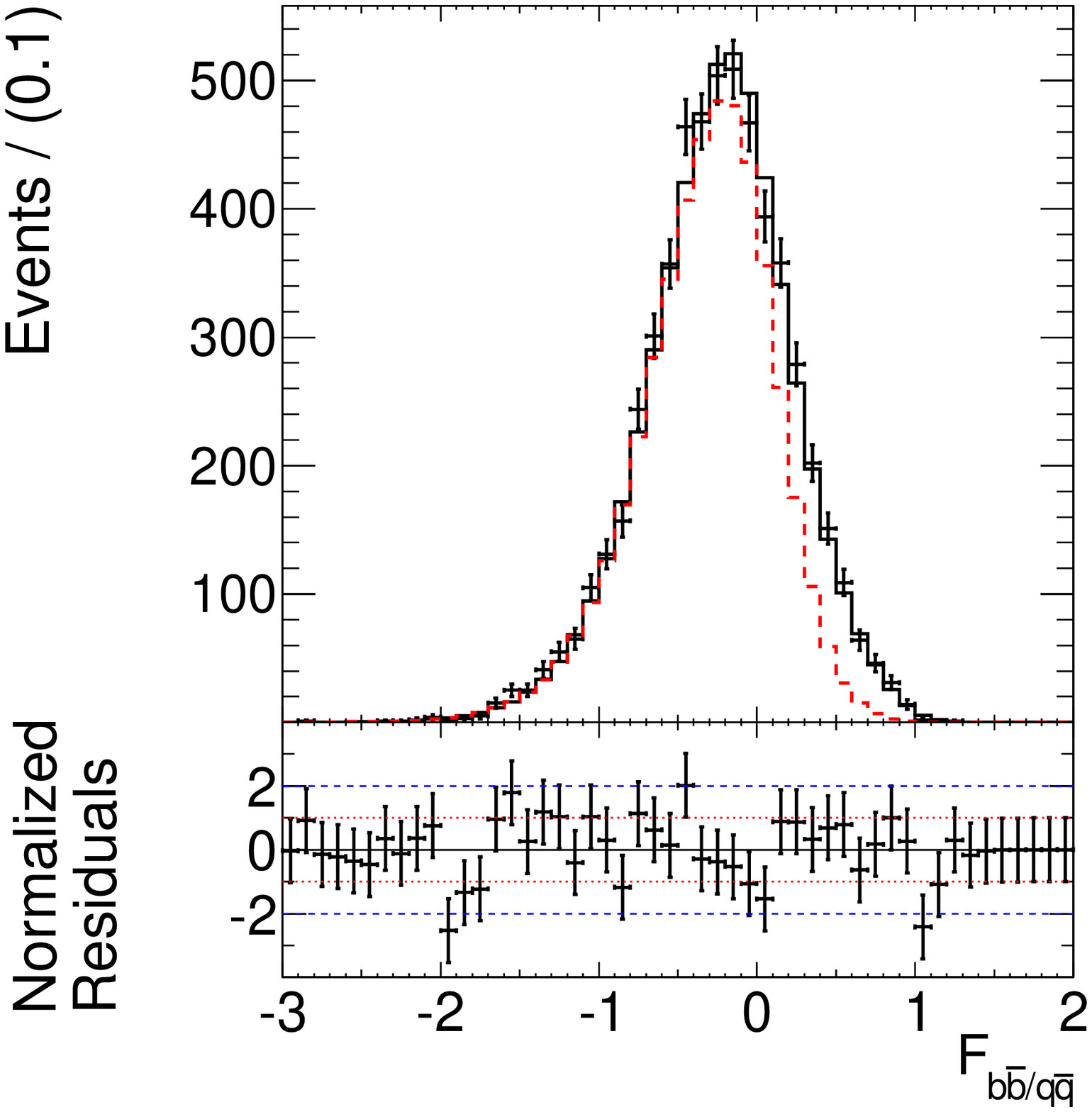}
  \put(-389,190){(a)}
  \put(-160,190){(b)}

  \includegraphics[height=220pt,width=!]{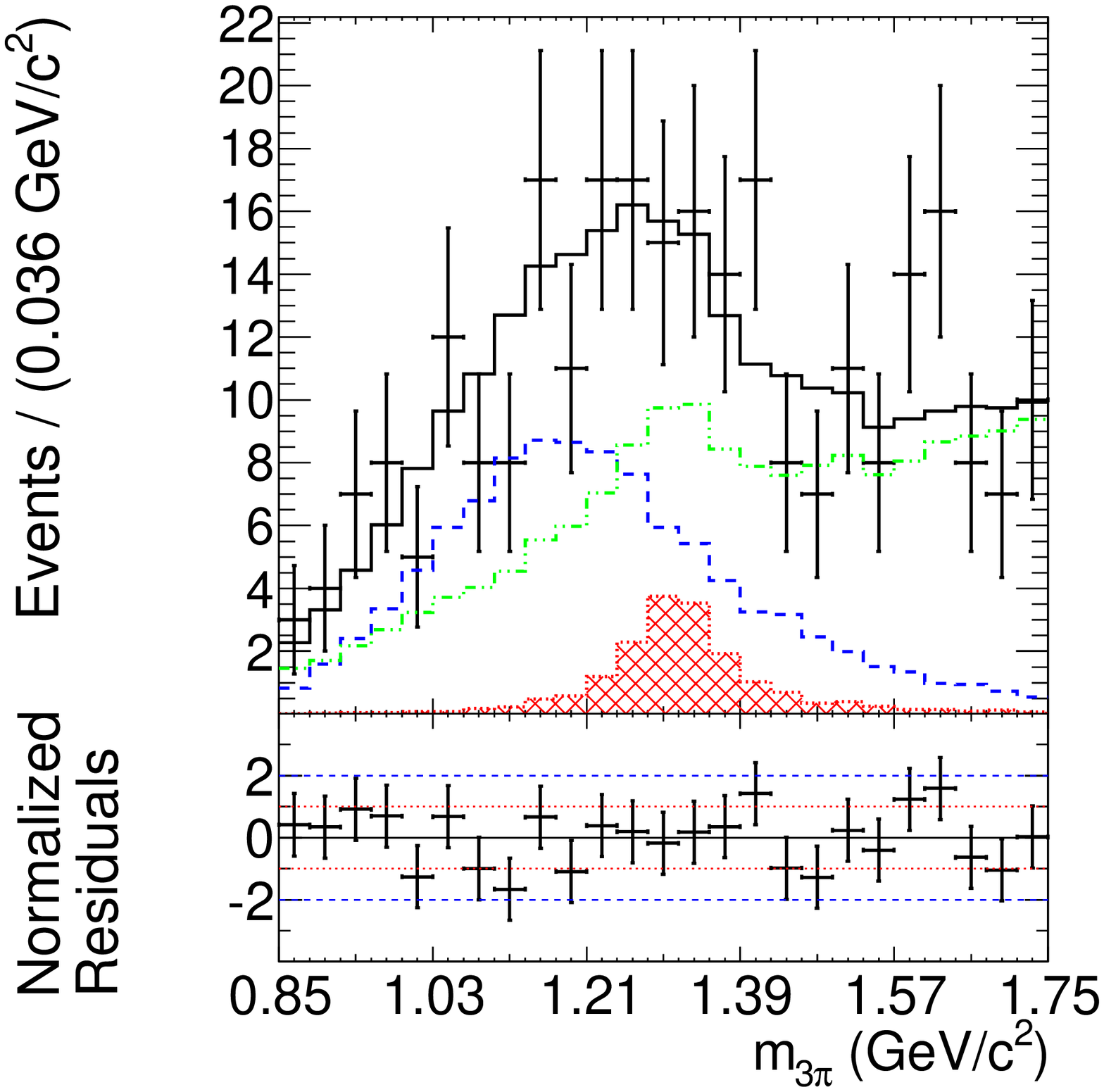}
  \includegraphics[height=220pt,width=!]{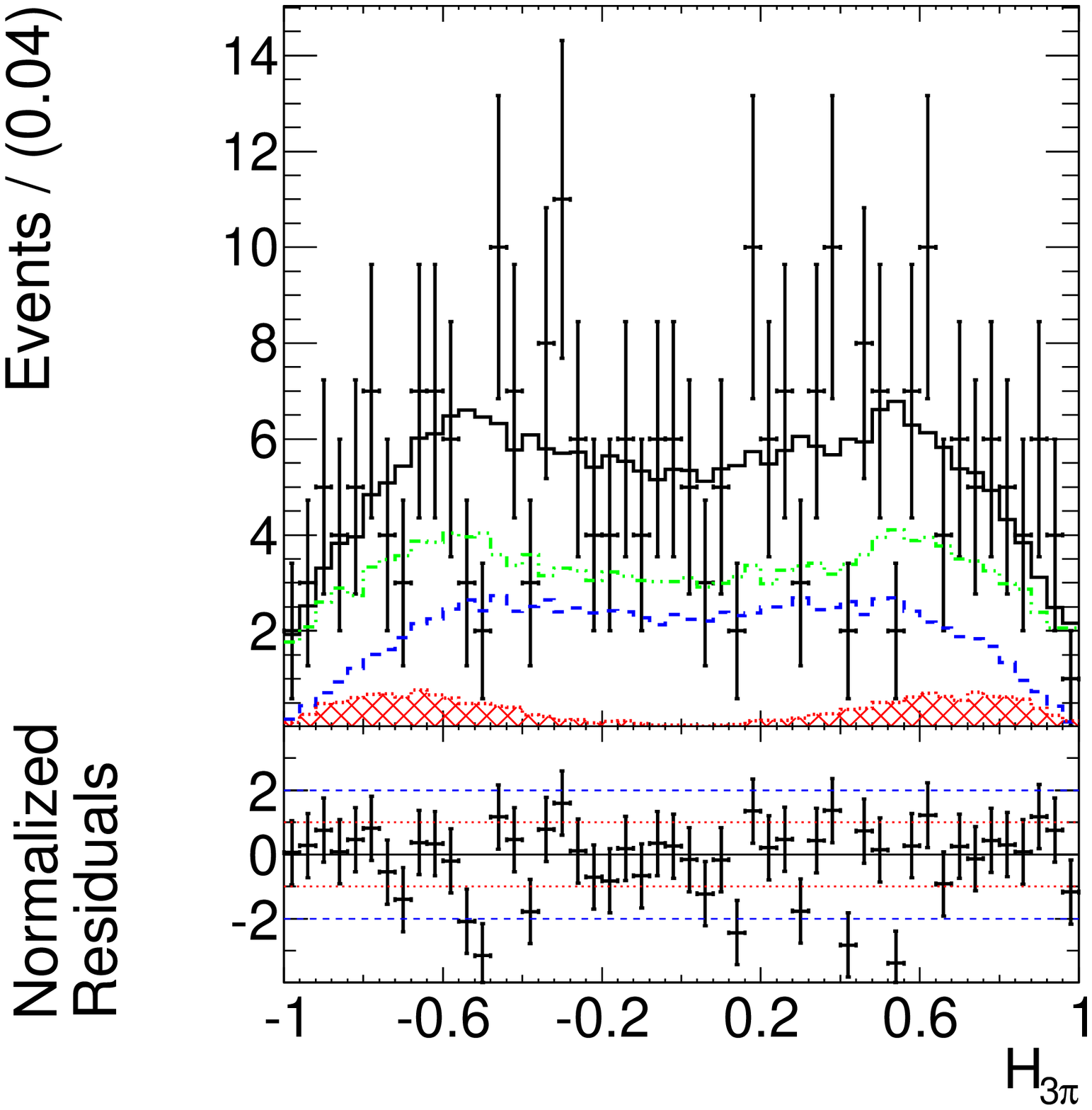}
  \put(-389,190){(c)}
  \put(-160,190){(d)}

  \caption{(color online) Projections of the fit to the \aonepi\ data. The points with error bars represent the data and the solid black histogram represents the fit result. (a) shows the \De\ projection for $\Fsb > 0.5$ and $r > 0.5$. The blue hatched histogram shows the peaking background, the dashed red histogram shows the continuum contribution and the dash-dotted green histogram shows the total background. (b) shows the \Fsb\ projection for $|\De| < 0.01 \; {\rm GeV}$ and $r > 0.5$. The dashed red histogram shows the continuum background contribution. (c) and (d) show the \mthreepi\ and \hel\ projections, respectively, for $\Fsb > 0.5$, $|\De| < 0.01 \; {\rm GeV}$ and $r > 0.5$. The dashed blue histogram shows the \aone\ contribution, the red hatched histogram shows the \atwo\ contribution and the dash-dotted green histogram shows the total background.}
  \label{fig_bf_data}
\end{figure}

We perform a fit to the data, with the projections shown in Fig.~\ref{fig_bf_data}, and obtain the product branching fractions
\begin{equation}
  {\cal B}(\aonepilong)\times{\cal B}(a^{\pm}_{1}(1260) \rightarrow \pi^{\pm}\pi^{\mp}\pi^{\pm}) = (11.1 \pm 1.0 \; (\rm stat) \pm 1.4 \; (\rm syst) ) \times 10^{-6},
\end{equation}
corresponding to a yield of $1445 \pm 216$ events, and
\begin{equation}
  {\cal B}(\atwopilong)\times{\cal B}(a^{\pm}_{2}(1320) \rightarrow \pi^{\pm}\pi^{\mp}\pi^{\pm})= (1.5 \pm 0.4 \; (\rm stat) \pm 0.4 \; (\rm syst) ) \times 10^{-6},
\end{equation}
corresponding to a yield of $282 \pm 106$ events. The statistical correlation coefficient between these two measurements is $-0.41$ and the statistical significance of the \aone\ peak is $16\sigma$, estimated by comparing the likelihood of the nominal fit result with that of a fit where the \aonepi\ branching fraction is fixed to zero. We also measure the \aone\ width to be $\Gamma_{a_{1}} = 381 \pm 43 \textrm{ (stat)}$ MeV.
From this fit, the relative contributions of each component are $0.8\%$ signal, $95.6\%$ continuum, $3.3\%$ \BBbar\ background and $0.3\%$ peaking background.

Our measurement of the \aonepi\ branching fraction is lower than that measured by the BaBar Collaboration~\cite{a1pi_BABAR1} by $1.9\sigma$ though still in marginal agreement. We are also in agreement with predictions made in the QCD factorization framework given in Refs.~\cite{theory_ap1,theory_ap2}, but not Ref.~\cite{theory_ap3}. The upper limit of the branching fraction of \atwopi\ is also determined to be
\begin{equation}
  {\cal B}(\atwopilong)\times{\cal B}(a^{\pm}_{2}(1320) \rightarrow \pi^{\pm}\pi^{\mp}\pi^{\pm}) < 2.2 \times 10^{-6} \textrm{ at 90\% CL},
\end{equation}
which is improved over the current world average~\cite{PDG} by about two orders of magnitude.

\subsection{Systematic Uncertainties}
Systematic errors from various sources are considered and estimated with independent internal studies and cross-checks. These are summarized in Table~\ref{tab_bf_syst}. This includes the uncertainty on the number of produced \BBbar\ events in the data sample. Contributions to the uncertainty in the selection efficiency due to particle identification and tracking are calculated by independent studies at Belle. The systematic uncertainty arising from the assumption that the $\aone$ decays exclusively through the dominant $\rhoz \pi^{\pm}$ intermediate state is accounted for by recalculating the detection efficiency with an exclusive $\fzsigma \pi^{\pm}$ MC.

The uncertainty in the \aone\ shape is determined by varying the fixed mass within its world average uncertainty~\cite{PDG}.
We account for a difference in the fraction of mis-reconstructed signal events between data and MC by varying this parameter by $\pm 5\%$ and repeating the fit.
Variations in the parametric model shape due to limited statistics are accounted for by varying each parameter within their errors. The dominant contribution to this category comes from the uncertainties in the signal shape correction factors obtained from analysing a high-statistics control sample \dpi.
Uncertainties in the non-parametric shapes are obtained by varying the contents of the histogram bins within $\pm 1\sigma$.
The systematic uncertainty due to fixing the peaking background yields is estimated by varying the branching fraction by its world average error~\cite{PDG} and repeating the fit. For modes where only an upper limit is known, the variation is taken as half of the upper limit.
The fit bias is determined from pseudo-experiments by searching for a difference between the generated and fitted physics parameters. As channels containing an $f_0$ are ignored in the nominal model, we account for a possible effect on the signal yield by embedding such events into these pseudo-experiments and determining further bias on the $a_1\pi$ and $a_2\pi$ branching fractions.
Finally, the uncertainty from neglecting interference between $a_1$ and $a_2$ was estimated by constructing a 4-body amplitude including detector effects and generating three relative interference configurations between $a_1$ and $a_2$: maximum constructive interference, no interference and maximal destructive interference. The largest deviation from the sample with no interference when fitting all with the nominal model gives the systematic uncertainty from interference.
\begin{table}
  \centering
  \caption{Systematic uncertainties of branching fractions.}
  \begin{tabular}
    {@{\hspace{0.5cm}}c@{\hspace{0.25cm}}  @{\hspace{0.25cm}}c@{\hspace{0.25cm}}  @{\hspace{0.25cm}}c@{\hspace{0.5cm}}}
    \hline \hline
    Category & $\delta{\cal B}(\aonepi)$ $(\%)$ & $\delta{\cal B}(\atwopi)$ $(\%)$\\
    \hline
    $N(\BBbar)$ & 1.4 & 1.4\\
    Tracking & 2.1 & 2.1\\
    Particle identification & 4.5 & 4.5\\
    $f_{0}(600)$ & 3.0 & N/A\\
    $a_{1}$ shape & 5.1 & 19.4\\
    Mis-Reconstruction fraction & 2.1 & 3.3\\
    Model shape & 4.2 & 14.0\\
    Histogram shape & 4.8 & 4.6\\
    Peaking background & 5.1 & 11.4\\
    Fit bias & 0.4 & 4.1\\
    Interference & 4.1 & 4.1\\\hline
    Total & 12.2 & 28.1\\
    \hline \hline
  \end{tabular}
  \label{tab_bf_syst}
\end{table}

\section{Time-dependent Measurement}
\label{Time-dependent Measurement}
\subsection{Event Selection}
In addition to the event selection criteria for the branching fraction measurement, events are selected for the time-dependent $CP$ violation measurement if they satisfy $|\De| < 0.04 \; {\rm GeV}$. This requirement retains $97\%$ of signal and $90\%$ of peaking backgrounds, while rejecting $60\%$ of the continuum background and $63\%$ of the \BBbar\ background. After this selection, the relative contributions of each component are $1.9\%$ signal, $94.2\%$ continuum, $3.1\%$ \BBbar\ background and $0.8\%$ peaking background.

\subsection{Event Model}
The signal PDF is given by
\begin{eqnarray}
  {\cal P}^{l,s}_{\rm Sig}(\Dt, q, c) &\equiv& (1+c\Acp)\frac{e^{-|\Dt|/\taub}}{8\taub} \biggl\{1-q\Dw^{l,s}+q(1-2w^{l,s})\times \nonumber \\
  & & \biggl[(\Scp + c\DS)\sin \Dmd \Dt - (\Ccp + c\DC)\cos \Dmd \Dt\biggr]\biggr\} \otimes R^{s}_{\BzBzb}(\Dt), \nonumber \\
\end{eqnarray}
which accounts for $CP$ dilution from the probability of incorrect flavor tagging $w^{l,s}$ and the wrong tag difference $\Dw^{l,s}$ between \Bz\ and \Bzb, both of which are determined from flavor specific control samples~\cite{Tagging}. This PDF is convolved with the \Dt\ resolution function for neutral $B$ particles $R^{s}_{\BzBzb}$, described in Ref.~\cite{ResFunc}.

The reconstructed vertex position of mis-reconstructed events is dominated by the high momentum bachelor pion from \aonepi. This pion is rarely mis-reconstructed, and the effect of borrowing a lower momentum track from the tag-side results in a slightly smaller lifetime. Thus, mis-reconstructed events are modelled using the truth model PDF with an effective lifetime and share $CP$ parameters with the truth model. The effective lifetime is determined from mis-reconstructed signal MC events and the effect of this choice is accounted for in the systematic uncertainties.

The continuum shape is determined from data taken below the \Ups\ resonance with the model
\begin{equation}
  P^{l,s}_{\qqbar}(\Dt, q, c) \equiv \frac{1 + qc\DC_{\qqbar}}{4} \biggl[(1 - f_{\delta}) \frac{e^{-|\Dt|/\tau_{\qqbar}}}{2\tau_{\qqbar}} + f_{\delta} \; \delta( \Dt - \mu^{s}_{\delta})\biggr] \otimes R^{s}_{\qqbar}(\Dt).
\end{equation}
This model contains a lifetime and prompt component to account for the charm and charmless contributions, respectively, and is convolved with a sum of two Gaussians
\begin{equation}
  R^{s}_{\qqbar}(\Dt) \equiv (1-f^{s}_{\rm tail})G(\Dt; \mu^{s}_{\rm mean}, S^{s}_{\rm main}\sigma) + f^{s}_{\rm tail}G(\Dt; \mu^{s}_{\rm mean}, S^{s}_{\rm main}S^{s}_{\rm tail}),
\end{equation}
which uses the event-dependent \Dt\ error constructed from the vertex resolution $\sigma \equiv (\sqrt{\sigma^{2}_{\rm Rec}+\sigma^{2}_{\rm Tag}})/\beta \gamma c$ as a scale factor. We also account for an asymmetry  in the product $qc$, with the parameter $\DC_{\qqbar}$, which is due to the jet-like topology of continuum. As a high momentum \pip(\pim) in \Brec\ is correlated with a high momentum \pim(\pip) on the tag-side, $q$ and $c$ will more often have the same sign.

The \BBbar\ shape is determined from a large MC sample and is divided into generic charmed and charmless, each further divided into neutral and charged \BBbar\ decays, using a lifetime model
\begin{equation}
  {\cal P}^{l,s}_{\BBbar}(\Dt, q, c) \equiv \frac{1 + cq\DC_{\BBbar}}{4}\frac{e^{-|\Dt|/\tau_{\BBbar}}}{2\tau_{\BBbar}}\otimes R^{s}_{\BBbar}(\Dt).
\end{equation}
where $R_{\BBbar}$ is relevant \Dt\ resolution function for either neutral or charged $B$ particles. Since reconstructed \BBbar\ events may borrow a particle from the tag side, the average \Dt\ lifetime tends to be smaller and is taken into account with the effective lifetime, $\tau_{\BBbar}$. Like continuum, \BBbar\ events also exhibit a $qc$ asymmetry; however, it is found to be more complex and is modelled with a 1st order polynomial in \Dt
\begin{equation}
  \DC_{\BBbar} \equiv p_{0} + p_{1}|\Dt|.
\end{equation}

The shapes of the peaking backgrounds are determined from individually generated MC events. As these backgrounds may exhibit $CP$ violation, we use a model similar to the truth model, but with an effective lifetime. We also fix all time-dependent parameters to null with the effects of this choice reflected in the systematic uncertainties.

To account for the broad underlying \Dt\ events not yet described by either signal or background PDFs, a broad Gaussian outlier PDF is introduced
\begin{equation}
  {\cal P}^{l,s}_{\rm Out}(\Dt, q, c) \equiv \frac{1}{4}G(\Dt; 0, \sigma^{s}_{\rm Out}).
\end{equation}

The total likelihood for $83799$ \aonepi\ candidates in the fit region becomes
\begin{equation}
  {\cal L} \equiv \prod_{l,s} \prod^{N_{l,s}}_{i=1} \sum_{j} f^{l,s}_{j}(\De^{i}, \Fsb^{i}, \mthreepi^{i}, \hel^{i}){\cal P}^{l,s}_{j}(\Dt^{i},q^{i}, c^{i})
\end{equation}
where $f^{l,s}_{j}$ is the event-dependent probability of component $j$, in flavor-tag bin $l$, with detector configuration $s$
\begin{equation}
  f^{l,s}_{j}(\De^{i}, \Fsb^{i}, \mthreepi^{i}, \hel^{i}) = \frac{N^{s}_{j}f^{l,s}_{j}{\cal P}^{l,s}_{j}(\De^{i}, \Fsb^{i}, \mthreepi^{i}, \hel^{i})}{\sum_{j}N^{s}_{j}f^{l,s}_{j}{\cal P}^{l,s}_{j}(\De^{i}, \Fsb^{i}, \mthreepi^{i}, \hel^{i})},
\end{equation}
constructed from the branching fraction measurement. Only the 5 time-dependent coefficients of the truth model are free in the fit to data.

As a consistency check, we perform a fit to data to measure the \Bz\ lifetime while fixing the 5 parameters of the truth model to zero. We obtain $\taub =  1.389 \pm 0.085 \; {\rm ps}$, which is in agreement with the current world average $\taub = 1.519 \pm 0.007 \; {\rm ps}$~\cite{PDG}.

\subsection{Fit Result}
\begin{figure}
  \centering
  \includegraphics[height=160pt,width=!]{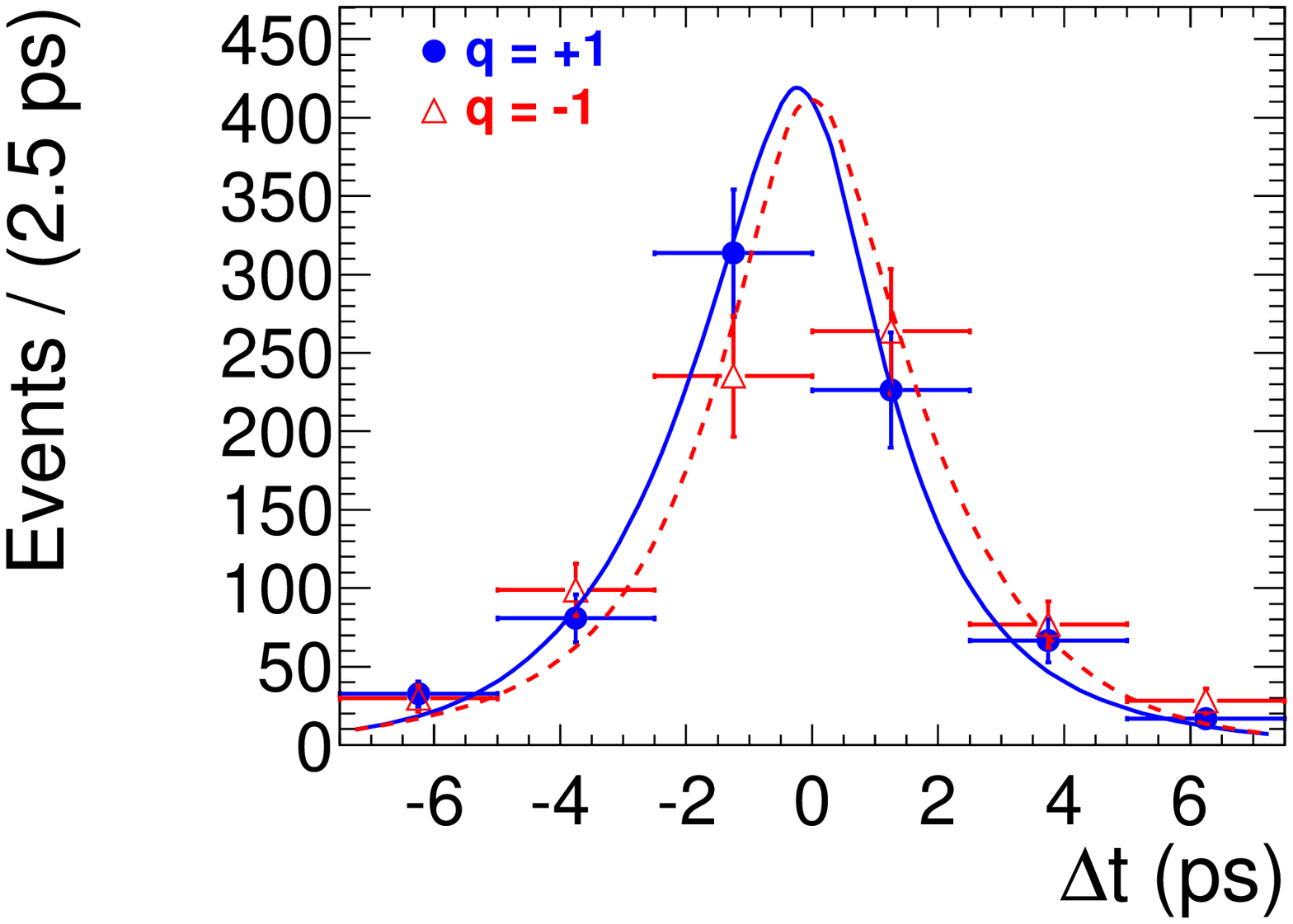}
  \includegraphics[height=160pt,width=!]{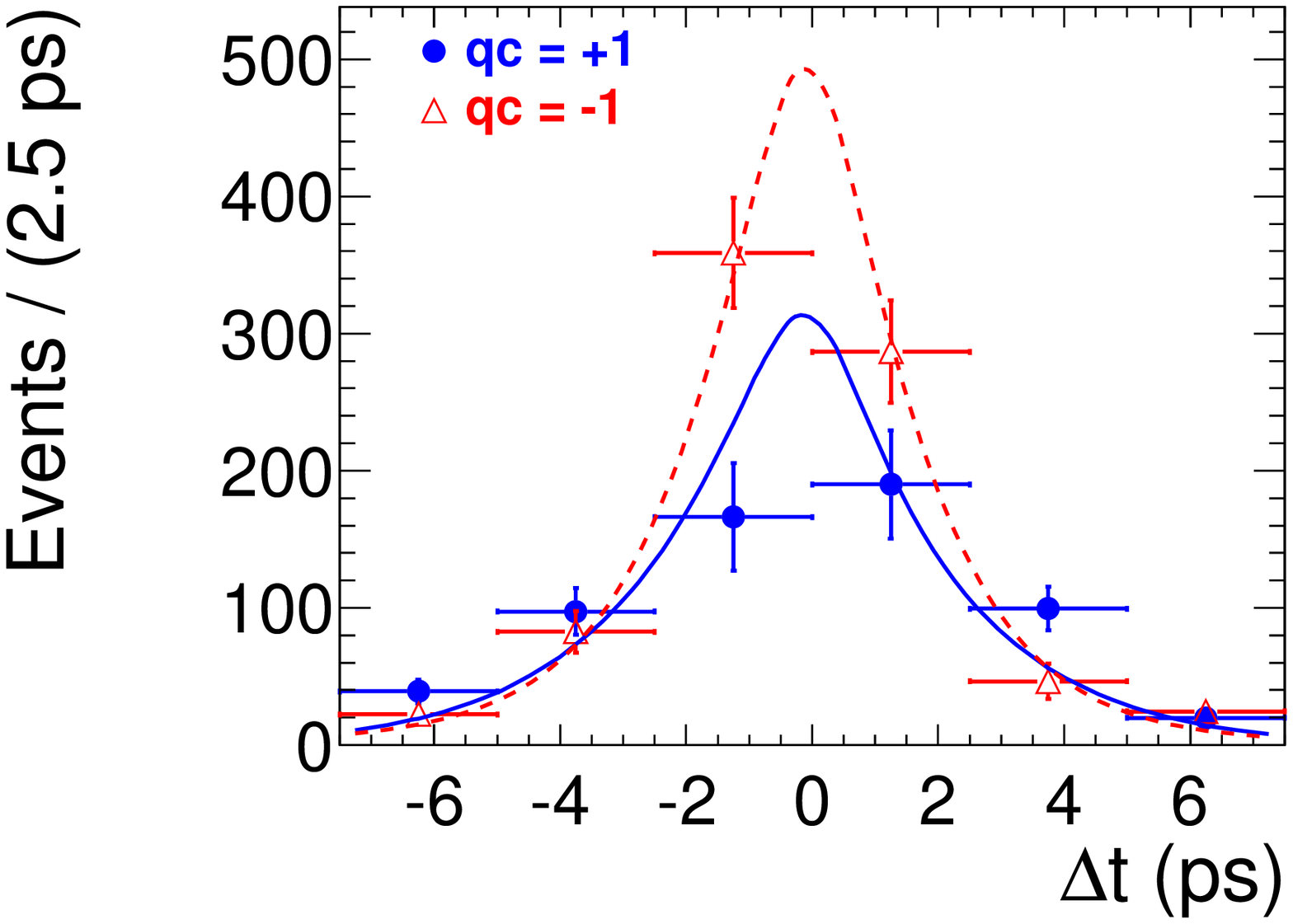}
  \put(-268,130){(a)}
  \put(-43,130){(b)}

  \includegraphics[height=160pt,width=!]{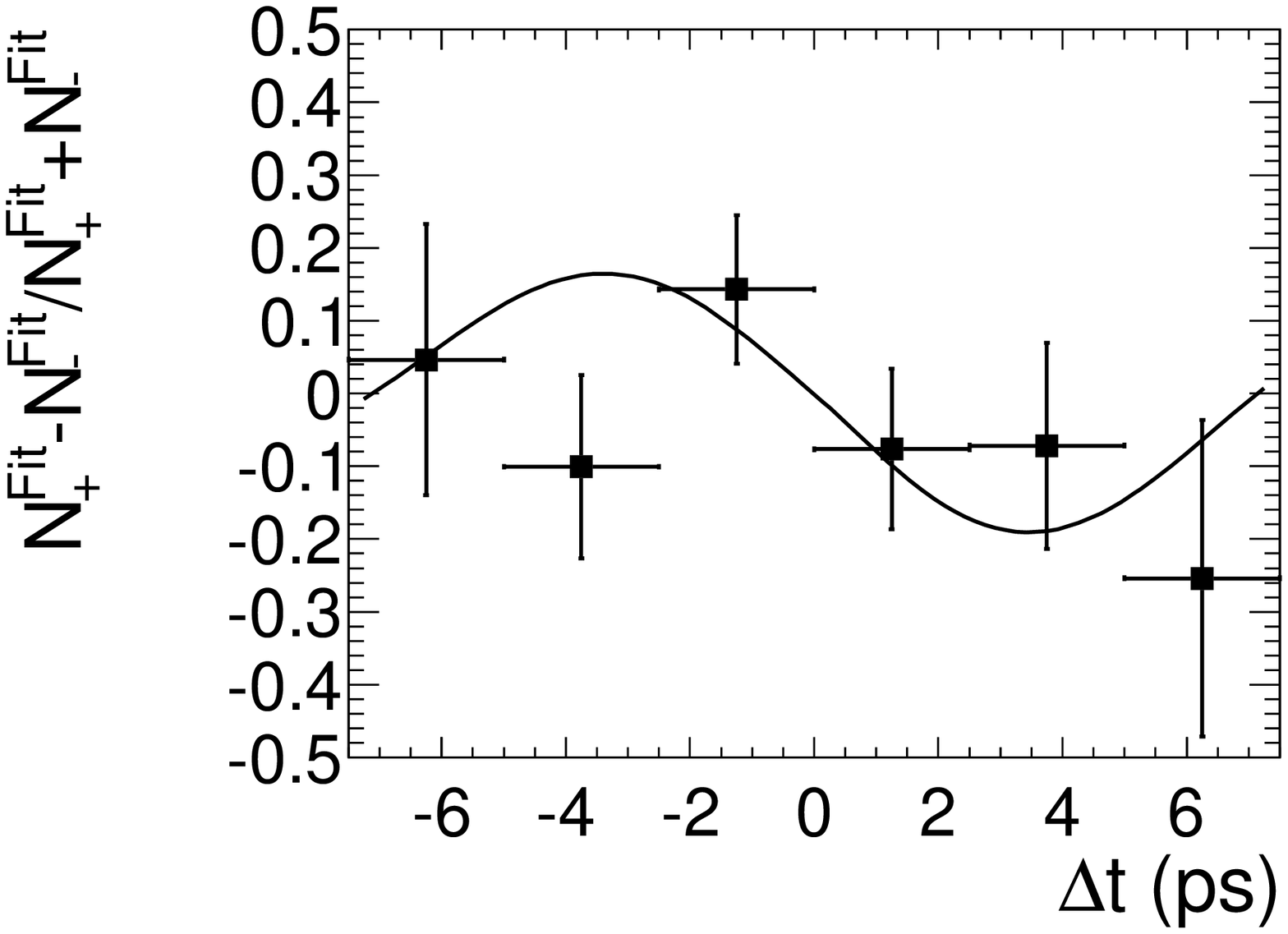}
  \includegraphics[height=160pt,width=!]{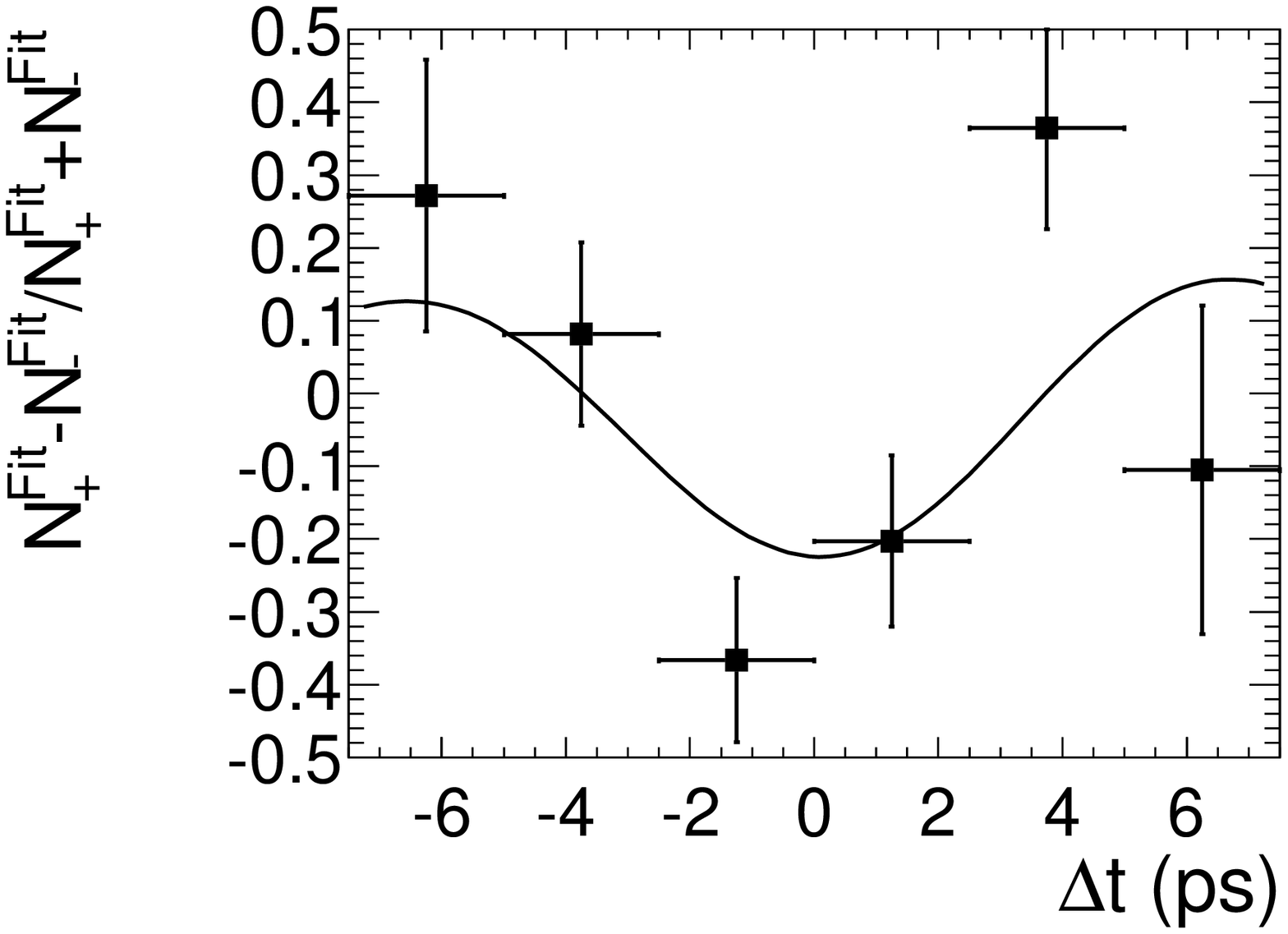}
  \put(-268,130){(c)}
  \put(-43,130){(d)}

  \caption{(color online) Background subtracted time-dependent fit results for \aonepi. (a) and (b) show the \Dt\ distributions for the \Btag\ flavor $q$ and the product of the \Btag\ flavor and $a_1$ charge $qc$, respectively. The left plots are useful for visualizing the effect of flavor-dependent $CP$ violation while the plots on the right show the effects of the $CP$ conserving parameters. The solid blue and dashed red curves represent the \Dt\ distributions for positive and negative quantity, respectively. (c) and (d) show the asymmetry of the plots immediately above them, $(N^{\rm Fit}_{+} - N^{\rm Fit}_{-})/(N^{\rm Fit}_{+} + N^{\rm Fit}_{-})$, where $N^{\rm Fit}_{+}$ ($N^{\rm Fit}_{-}$) is the measured signal yield of positive (negative) quantities in bins of \Dt.}
  \label{fig_cp_tcpv}
\end{figure}

We perform a fit to the data and obtain the $CP$ violating parameters
\begin{equation}
  \begin{array}{rcl}
    \Acp \!&=&\! -0.06 \pm 0.05 \textrm{ (stat)} \pm 0.07 \textrm{ (syst)},\\
    \Ccp \!&=&\! -0.01 \pm 0.11 \textrm{ (stat)} \pm 0.09 \textrm{ (syst)},\\
    \Scp \!&=&\! -0.51 \pm 0.14 \textrm{ (stat)} \pm 0.08 \textrm{ (syst)},
  \end{array}
\end{equation}
and the $CP$ conserving parameters
\begin{equation}
  \begin{array}{rcl}
    \DC \!&=&\! +0.54 \pm 0.11 \textrm{ (stat)} \pm 0.07 \textrm{ (syst)},\\
    \DS \!&=&\! -0.09 \pm 0.14 \textrm{ (stat)} \pm 0.06 \textrm{ (syst)}.
  \end{array}
\end{equation}
The background subtracted fit results are shown in Fig.~\ref{fig_cp_tcpv}, where the data points are the signal yields obtained by repeating the branching fraction fits in $\Dt, q$ or $\Dt, qc$ bins, accounting for the selection criteria on \De.
Ours are the most precise measurements of these parameters to date and are in agreement with those obtained by the BaBar Collaboration~\cite{a1pi_BABAR2}. We are also in very good agreement with the theoretical predictions made in Ref.~\cite{theory_ap1}. The statistical correlation coefficients between the obtained parameters are given in Table~\ref{tab_cp_corr}.
\begin{table}
  \centering
  \caption{Statistical correlation matrix for the fit result.}
  \begin{tabular}
    {@{\hspace{0.5cm}}c@{\hspace{0.25cm}}  @{\hspace{0.25cm}}c@{\hspace{0.25cm}}  @{\hspace{0.25cm}}c@{\hspace{0.25cm}}  @{\hspace{0.25cm}}c@{\hspace{0.25cm}}  @{\hspace{0.25cm}}c@{\hspace{0.25cm}}  @{\hspace{0.25cm}}c@{\hspace{0.5cm}}}
    \hline \hline
    & \Acp & \Ccp & \DC & \Scp & \DS\\
    \hline
    \Acp & $1$ & & & & \\
    \Ccp & $-0.20$ & $1$ & & & \\
    \DC & $+0.01$ & $+0.03$ & $1$ & & \\
    \Scp & $+0.02$ & $-0.01$ & $-0.03$ & $1$ & \\
    \DS & $+0.09$ & $-0.05$ & $-0.01$ & $+0.02$ & $1$\\
    \hline \hline
  \end{tabular}
  \label{tab_cp_corr}
\end{table}

A MC technique is employed to obtain \phitwoeff\ using Eq.~\ref{eq_phitwo} in order to take correlations between the fitted parameters into account. We generate multiple vectors $(\Ccp, \DC, \Scp, \DS)$ based on a correlated multi-dimensional Gaussian constructed from the fit result and solve for the four solutions of \phitwoeff\ each time. We take the central values and their uncertaintes from the resulting distributions of \phitwoeff\ and obtain the four solutions
\begin{align}
  \phitwoeff &= (-17.3 \pm 6.6 \textrm{ (stat)} \pm 4.8 \textrm{ (syst)})^{\circ},\nonumber\\
  \phitwoeff &= (41.6 \pm 6.2 \textrm{ (stat)} \pm 3.4 \textrm{ (syst)})^{\circ},\nonumber\\
  \phitwoeff &= (48.4 \pm 6.2 \textrm{ (stat)} \pm 3.4 \textrm{ (syst)})^{\circ},\nonumber\\
  \phitwoeff &= (107.3 \pm 6.6 \textrm{ (stat)} \pm 4.8 \textrm{ (syst)})^{\circ}.
\end{align}
Using a similar technique, we obtain the direct $CP$ violation parameters given in Eq.~\ref{eq_apm}
\begin{align}
  A_{+-} &= +0.07 \pm 0.08 \textrm{ (stat)} \pm 0.10 \textrm{ (syst)},\nonumber\\
  A_{-+} &= -0.04 \pm 0.26 \textrm{ (stat)} \pm 0.19 \textrm{ (syst)},
\end{align}
where the statistical correlation coefficient between these two parameters is 0.61.

\begin{figure}
  \centering
  \includegraphics[height=160pt,width=!]{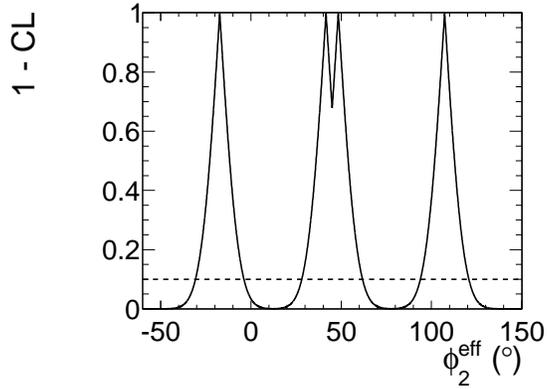}
  \caption{Difference 1-CL plotted for a range of \phitwoeff\ values as shown by the solid curve. The $p$-value is taken as the maximum $p$-value that can be calculated from all four solutions. The dashed line indicates CL $= 90\%$.}
  \label{fig_CL}
\end{figure}

\begin{figure}
  \centering
  \includegraphics[height=160pt,width=!]{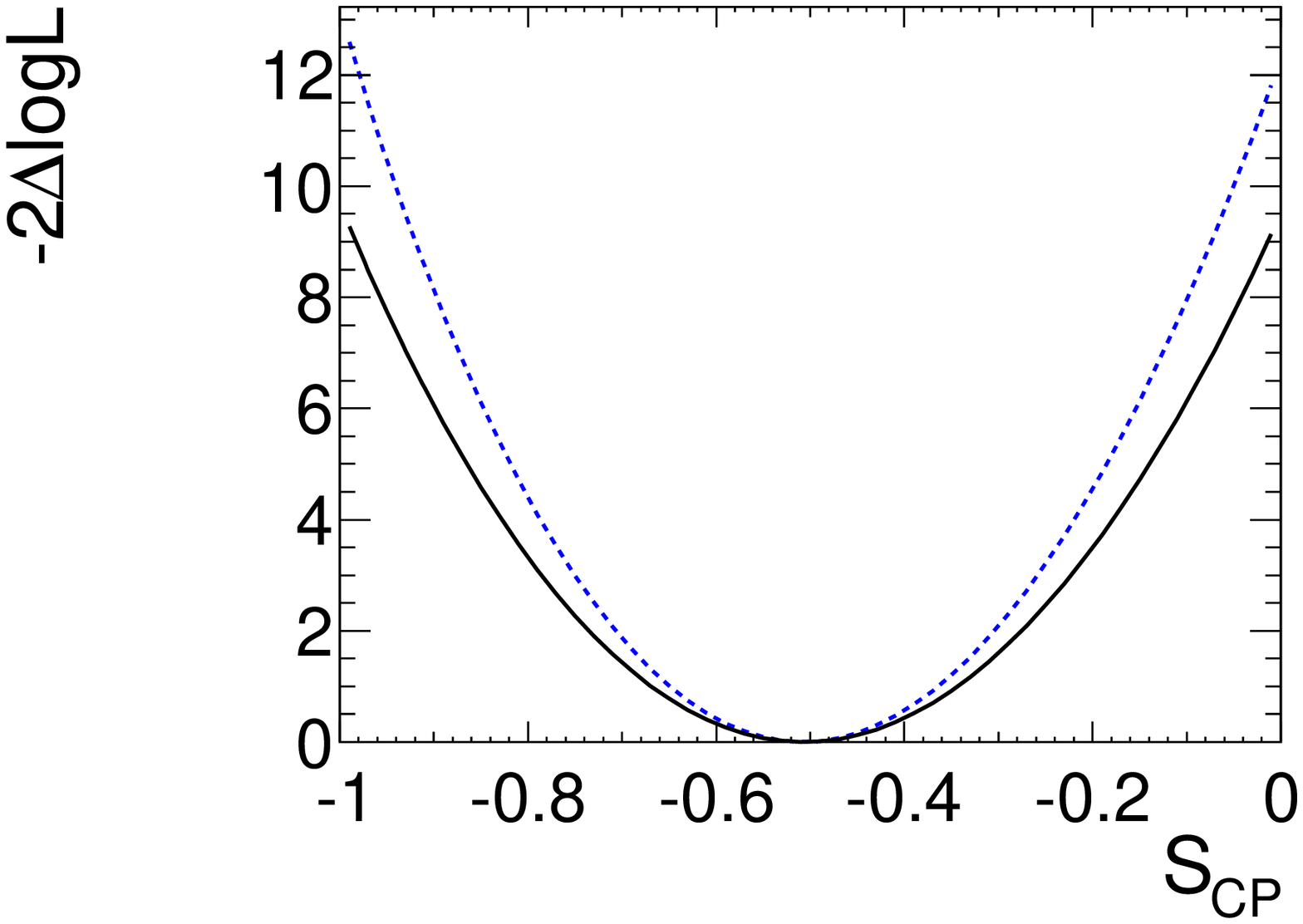}
  \includegraphics[height=160pt,width=!]{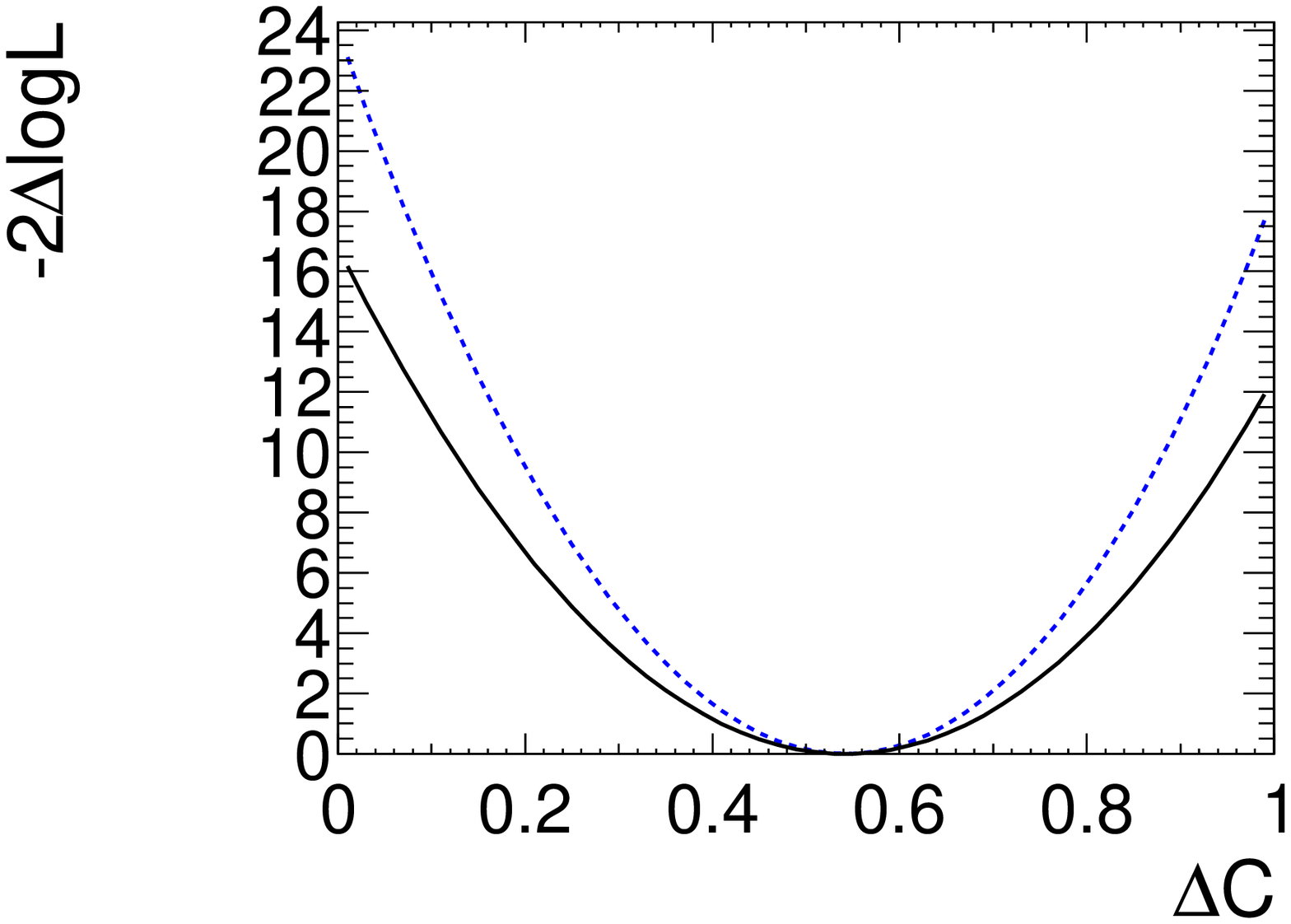}
  \put(-268,130){(a)}
  \put(-43,130){(b)}

  \caption{(color online) Likelihood scan of \Scp\ (a) and \DC\ (b). The dotted blue curve shows the statistical likelihood while the solid black curve includes the systematic uncertainty.}
  \label{fig_cp_sigma}
\end{figure}

We compose Gaussian distributions from the four solutions for \phitwoeff\ and construct a two-sided $p$-value plot for \phitwoeff\ as shown in Fig.~\ref{fig_CL}.
We also perform a likelihood scan to estimate the significance of \Scp\ and \DC\ as shown in Fig.~\ref{fig_cp_sigma}. The significance of mixing-induced $CP$ violation is found to be $3.1\sigma$ including systematic uncertainties while the rate where the \aone\ does not contain the spectator quark is found to dominate the rate where it does at the $4.1\sigma$ level.

\subsection{Systematic Uncertainties}
The systematic uncertainties for the time-dependent parameters are summarized in Table~\ref{tab_cp_syst}. These are estimated from various sources including uncertainties in the IP profile, charged track selection based on track helix errors, helix parameter corrections, \Dt\ and vertex goodness-of-fit selection, \Dz\ bias and SVD misalignment. The fixed physics parameters \taub\ and \Dmd, \Dt\ resolution function and data model shape parameters including background effective lifetimes and $qc$ asymmetries, as well as the flavor tagging performance parameters $w$ and \Dw, are varied by $\pm 1 \sigma$. We generate MC pseudo-experiments and perform an ensemble test to obtain systematic biases from interference on the tag-side arising between the CKM favored $b \bar d \to (c \bar u d) \bar d$ and doubly-CKM-suppressed $\bar b d \to (\bar u c \bar d) d$ amplitudes in final states used for flavor tagging~\cite{tsi}. These sources should not affect \Acp, as this parameter represents a time and flavor-integrated asymmetry. 

The remaining systematic categories affect all time-dependent parameters. The parameters and non-parametric shapes describing signal probability are varied in the same way as for the branching fraction measurement. The $CP$ violation parameters of mis-reconstructed events are assumed to be the same as signal. To account for this, a sample of GEANT MC is produced with the nominal fit result. The systematic error is taken as the difference between the fit result to the correctly reconstructed subsample and a fit to the whole sample sharing the $CP$ parameters between the signal and mis-reconstructed components. The fit bias is determined from an ensemble test by searching for a difference between the generated and fitted physics parameters.

Possible $CP$ violation in the background is the dominant systematic uncertainty. We assume that the neutral \BBbar\ background possesses a 20\% $CP$ violating component and that peaking backgrounds display a 50\% $CP$ violating effect. For the neutral \BBbar\ component, the uncertainty is determined by fixing the $CP$ parameters and refitting the data. For the peaking backgrounds, special GEANT MC samples are produced with the nominal fit result for signal and $CP$ violation generated in the peaking background. We perform a fit with the peaking background $CP$ parameters fixed to null asymmetry and compare this with a fit where they are fixed to the generated values.

To estimate the effects of interference on the signal probability, we employ the same 4-body amplitude generator described in the branching fraction measurement. MC samples are generated with the nominal fit result with the three phase configurations: maximum constructive interference, no interference and maximal destructive interference. Firstly, signal probability is predetermined for all three samples by fitting for the branching fractions of the $a_1\pi$ and $a_2\pi$ components assuming no interference in the fit model. Finally, a time-dependent fit is performed to the three phase configurations and the systematic error taken as the maximum deviation from the sample with no generated interference.
\begin{table}
  \footnotesize
  \centering
  \caption{Systematic uncertainties of time-dependent parameters. Categories related to \Dt\ reconstruction and flavor tagging are not applicable to the time and flavor integrated \Acp.}
  \begin{tabular}
    {@{\hspace{0.5cm}}c@{\hspace{0.25cm}}  @{\hspace{0.25cm}}c@{\hspace{0.25cm}}  @{\hspace{0.25cm}}c@{\hspace{0.25cm}}  @{\hspace{0.25cm}}c@{\hspace{0.25cm}}  @{\hspace{0.25cm}}c@{\hspace{0.25cm}}  @{\hspace{0.25cm}}c@{\hspace{0.5cm}}}
    \hline \hline
    Category & $\delta\Acp$ $(10^{-2})$ & $\delta\Ccp$ $(10^{-2})$ & $\delta\DC$ $(10^{-2})$ & $\delta\Scp$ $(10^{-2})$ & $\delta\DS$ $(10^{-2})$\\
    \hline
    IP profile & N/A & 0.2 & 0.2 & 1.0 & 1.0\\
    \Btag\ track selection & N/A & 1.2 & 0.4 & 0.8 & 1.1\\
    Track helix errors & N/A & 0.0 & 0.0 & 0.0 & 0.0\\
    \Dt\ selection & N/A & 0.1 & 0.0 & 0.1 & 0.0\\
    Vertex quality selection & N/A & 0.3 & 0.9 & 0.3 & 0.2\\
    $\Delta z$ bias & N/A & 0.5 & 0.5 & 0.4 & 0.4\\
    Misalignment & N/A & 0.4 & 0.4 & 0.2 & 0.2\\
    \taub\ and \Dmd & N/A & 0.3 & 0.3 & 0.2 & 0.2\\
    \Dt\ resolution function & N/A & 1.3 & 0.9 & 2.8 & 1.7\\
    Flavor tagging & N/A & 0.3 & 0.2 & 0.2 & 0.1\\
    Model shape & N/A & 0.3 & 2.9 & 0.6 & 0.5\\
    Tag-side interference & N/A & 3.6 & 0.2 & 0.5 & 0.4\\
    Signal probability & 0.5 & 0.4 & 1.9 & 2.0 & 0.8\\
    Mis-Reconstruction & 0.2 & 0.1 & 0.7 & 0.3 & 0.3\\
    Fit bias & 0.4 & 1.6 & 0.3 & 1.0 & 0.1\\
    Background $CP$ violation & 6.6 & 7.6 & 5.7 & 6.8 & 5.1\\
    Interference & 0.8 & 1.1 & 2.0 & 0.2 & 1.0\\\hline
    Total & 6.6 & 8.9 & 7.2 & 7.9 & 5.8\\
    \hline \hline
  \end{tabular}
  \label{tab_cp_syst}
\end{table}

\section{Conclusion}
\label{Conclusion}
We have presented a measurement of the product branching fraction and time-dependent parameters in \aonepilong\ decays, which are in agreement with measurements performed by the BaBar Collaboration~\cite{a1pi_BABAR1,a1pi_BABAR2}. We obtain the product branching fraction
\begin{equation}
  {\cal B}(\aonepilong)\times{\cal B}(a^{\pm}_{1}(1260) \rightarrow \pi^{\pm}\pi^{\mp}\pi^{\pm}) = (11.1 \pm 1.0 \; (\rm stat) \pm 1.4 \; (\rm syst) ) \times 10^{-6},
\end{equation}
and an upper limit on the product branching fraction for a possible decay with the same final state
\begin{equation}
  {\cal B}(\atwopilong)\times{\cal B}(a^{\pm}_{2}(1320) \rightarrow \pi^{\pm}\pi^{\mp}\pi^{\pm}) < 2.2 \times 10^{-6} \textrm{ at 90\% CL}.
\end{equation}
This upper limit is an improvement over the current world's most restrictive limit by about two orders of magnitude. In a time-dependent measurement to extract $CP$ asymmetries, we obtain the $CP$ violation parameters
\begin{equation}
  \begin{array}{rcl}
    \Acp \!&=&\! -0.06 \pm 0.05 \textrm{ (stat)} \pm 0.07 \textrm{ (syst)},\\
    \Ccp \!&=&\! -0.01 \pm 0.11 \textrm{ (stat)} \pm 0.09 \textrm{ (syst)},\\
    \Scp \!&=&\! -0.51 \pm 0.14 \textrm{ (stat)} \pm 0.08 \textrm{ (syst)},
  \end{array}
\end{equation}
representing time and flavor integrated direct, flavor-dependent direct and mixing-induced $CP$ violation, respectively. Simultaneously, we also extract the $CP$ conserving parameters
\begin{equation}
  \begin{array}{rcl}
    \DC \!&=&\! +0.54 \pm 0.11 \textrm{ (stat)} \pm 0.07 \textrm{ (syst)},\\
    \DS \!&=&\! -0.09 \pm 0.14 \textrm{ (stat)} \pm 0.06 \textrm{ (syst)},
  \end{array}
\end{equation}
which, respectively, describe a rate difference and strong phase difference between the decay channels where the \aone\ does not contain the spectator quark and those where it does. We find first evidence of mixing-induced $CP$ violation in \aonepilong\ decays with $3.1\sigma$ significance and the rate where the \aone\ does not contain the spectator quark is found to dominate the rate where it does at the $4.1\sigma$ level. However, there is no evidence for either time and flavor integrated direct $CP$ violation or flavor-dependent direct $CP$ violation. Our results are in good agreement with theoretical predictions given within the QCD factorization framework~\cite{theory_ap1,theory_ap2} and may be used in an 
either an isospin analysis~\cite{theory_isospin} or $SU(3)$ flavor symmetry~\cite{theory_a1pi} to extract \phitwo.

\section*{ACKNOWLEDGMENTS}
%***** Acknowledgments *****

We thank the KEKB group for the excellent operation of the
accelerator; the KEK cryogenics group for the efficient
operation of the solenoid; and the KEK computer group,
the National Institute of Informatics, and the 
PNNL/EMSL computing group for valuable computing
and SINET4 network support.  We acknowledge support from
the Ministry of Education, Culture, Sports, Science, and
Technology (MEXT) of Japan, the Japan Society for the 
Promotion of Science (JSPS), and the Tau-Lepton Physics 
Research Center of Nagoya University; 
the Australian Research Council and the Australian 
Department of Industry, Innovation, Science and Research;
the National Natural Science Foundation of China under
contract No.~10575109, 10775142, 10875115 and 10825524; 
the Ministry of Education, Youth and Sports of the Czech 
Republic under contract No.~LA10033 and MSM0021620859;
the Department of Science and Technology of India; 
the Istituto Nazionale di Fisica Nucleare of Italy; 
the BK21 and WCU program of the Ministry Education Science and
Technology, National Research Foundation of Korea,
and GSDC of the Korea Institute of Science and Technology Information;
the Polish Ministry of Science and Higher Education;
the Ministry of Education and Science of the Russian
Federation and the Russian Federal Agency for Atomic Energy;
the Slovenian Research Agency;  the Swiss
National Science Foundation; the National Science Council
and the Ministry of Education of Taiwan; and the U.S.\
Department of Energy and the National Science Foundation.
This work is supported by a Grant-in-Aid from MEXT for 
Science Research in a Priority Area (``New Development of 
Flavor Physics''), and from JSPS for Creative Scientific 
Research (``Evolution of Tau-lepton Physics'').
We also thank our theory colleagues at the Max-Planck-Institut f\"{u}r Physik, S.~Borowka and W.~Ochs, for helpful discussions.

\end{document}